\documentclass[twocolumn]{IEEEtran} 
\usepackage[final]{graphicx}
\usepackage{subfigure}
\usepackage{footnote}
\makesavenoteenv{tabular}
\usepackage{threeparttable}
\usepackage[cmex10]{amsmath}
\usepackage{cases}
\usepackage{multirow}

\def\BibTeX{{\rm B\kern-.05em{\sc i\kern-.025em b}\kern-.08em
    T\kern-.1667em\lower.7ex\hbox{E}\kern-.125emX}}

\begin{document}

\title{Simultaneous Transmission and Reception: Algorithm, Design and System Level Performance} 

\author{Yang-Seok ~Choi, ~\IEEEmembership{Member, ~IEEE,} and Hooman
~Shirani-Mehr,~\IEEEmembership{Member, ~IEEE}%
\thanks{Copyright (c) 2013 IEEE. Personal use of this material is permitted.
However, permission to use this material for any other purposes must
be obtained from the IEEE by sending a request to
pubs-permissions@ieee.org.}
\thanks{Manuscript received August 7
2012; revised December 20 2012, May 14 2013, and August 15 2013;
accepted September 16 2013. The associate editor coordinating the
review of this paper and approving it for publication was Prof. G. Xing.}%

\thanks{Y.-S. Choi was with Mobile and Communications Group, Intel
Corporation, Hillsboro, OR 97124, USA. He is now with Intel Labs,
Intel Corporation, Hillsboro, OR 97124, USA (email:
yang-seok.choi@intel.com).}%
\thanks{ H. Shirani-Mehr is with Mobile
and Communications Group, Intel Corporation, Hillsboro, OR 97124,
USA (email: hooman.shirani-mehr@intel.com).} }

\markboth{To Appear in IEEE Transactions On Wireless Communications}{Choi, \MakeLowercase{\textit{et al.}}: Simultaneous Transmission and Reception: Algorithm, Design and System Level Performance} 

\maketitle

\begin{abstract}
Full Duplex or Simultaneous transmission and reception (STR) in the
same frequency at the same time can potentially double the physical
layer capacity. However, high power transmit signal will appear at
receive chain as echoes with powers much higher than the desired
received signal. Therefore, in order to achieve the potential gain,
it is imperative to cancel these echoes. As these high power echoes
can saturate low noise amplifier (LNA) and also digital domain echo
cancellation requires unrealistically high resolution
analog-to-digital converter (ADC), the echoes should be cancelled or
suppressed sufficiently before LNA. In this paper we present a
closed-loop echo cancellation technique which can be implemented
purely in analogue domain. The advantages of our method are
multiple-fold: it is robust to phase noise, does not require
additional set of antennas, can be applied to wideband signals and
the performance is irrelevant to radio frequency (RF) impairments in
transmit chain. Next, we study a few protocols for STR systems in
carrier sense multiple access (CSMA) network and investigate MAC
level throughput with realistic assumptions in both single cell and
multiple cells. We show that STR can reduce hidden node problem in
CSMA network and produce gains of up to 279\% in maximum throughput
in such networks. Moreover, at high traffic load, the gain of STR
system can be tremendously large since the throughput of non-STR
system is close to zero at heavy traffic due to severe collisions.
Finally, we investigate the application of STR in cellular systems
and study two new unique interferences introduced to the system due
to STR, namely \textit{BS-BS interference} and \textit{UE-UE
interference}. We show that these two new interferences will hugely
degrade system performance if not treated appropriately. We propose
novel methods to reduce both interferences and investigate the
performances in system level. We show that BS-BS interference can be
suppressed sufficiently enough to be less than thermal noise power,
and with favorable UE-UE channel model, capacities close to double
are observed both in downlink (DL) and uplink (UL). When UE-UE
interference is larger than DL co-channel interferences, we propose
a simple and ``non-cooperative" technique in order to reduce UE-UE
interference.

\end{abstract}

\begin{keywords}
Full duplex, Simultaneous Transmission and Reception (STR), Echo
cancellation, CSMA, Null forming, Hidden node.
\end{keywords}

\section{Introduction}
\PARstart{C}{u}rrently deployed wired and wireless communication
systems employ half duplex. Namely, either frequency division duplex
(FDD) or time division duplex (TDD) has been used for separate
transmission and reception. In FDD and TDD, transmitted signal does
not interfere with received signal due to orthogonal use of
time/frequency resources for transmission and reception. Since two
orthogonal channels are needed in half duplex systems, twice of time
and/or frequency resources are required in half duplex compared to
full duplex systems. It is clear that the capacity can be doubled by
simultaneous transmission and reception in the same frequency at the
same time.

\subsection{Benefits of Full Duplex}

STR or full duplex systems not only improve the physical layer
capacity but also provide other important benefits in layers beyond
physical layer \cite{stanford}. For example, STR can reduce or
completely eliminate hidden node problem which is typical issue in
CSMA networks such as wireless local area networks (WLAN). When a
node receives a packet designated to it and meanwhile has a packet
to transmit, by having STR capability it can transmit the packet
while receiving the designated packet. This not only provides twice
throughput but also enables hidden nodes to better detect active
nodes in their neighborhoods. On the other hand, when the node has
no packet to send, it can transmit a dummy signal so that any hidden
node can detect the activity in its vicinity and realize that the
channel is in use.

Another benefit of STR is significant reduction in end-to-end delay
in multi-hop networks \cite{stanford}. In half duplex systems, each
node can start transmission of a packet to the next node only when
it is fully received from the previous node in network. Therefore,
the end-to-end delay is equal to packet duration multiplied by the
number of hops. However, when STR is employed, a node can forward a
packet while receiving it, and consequently the end-to-end delay in
STR systems can be just a bit longer than the packet duration. This
will be a huge advantage over half duplex systems especially as the
number of hops grows. Meanwhile, the forwarded packet to next node
can play a role of implicit acknowledgement (ACK) to previous node
as well.

Interesting application of STR includes channel sensing in cognitive
radio systems \cite{stanford}. In cognitive radios, active secondary
users have to release the spectrum when primary users start their
transmissions. Without STR capability, it is a challenge for
secondary users to detect activity of primary users while they are
using the spectrum for their own communications. However, STR
enabled secondary users can scan the activities of primary users
frequently (even as they transmit) and stop their transmissions
immediately once they detect primary users' transmissions.

Likewise, STR makes the device discovery easier in device-to-device
(D2D) systems. This is due to the fact that in D2D systems, when
user equipment (UE) has STR capability, it can discover neighboring
UEs easily by monitoring UL signals from proximate UEs without
stopping its own UL transmission.

It is interesting to note that STR techniques can be used for
interference cancellation in co-existence of multiple radios in the
same device. Multiple radios such as WLAN, Bluetooth, GPS receiver,
2G and 3G cellular transceivers are put into the same device
particularly a small handheld device type \cite{mac3}. Although
those radios operate at different RF carriers, due to proximity of
transceivers in the same device, they can still interfere with each
other. This interference can be treated as echo since the
transmitted signal from a radio is arrived at other radios in the
same device. Hence, by using the proposed technique in this paper,
the co-existence issues in the same device can be resolved.

Multiband support requires large number of switched duplexers in
FDD, resulting in quite complicated RF architecture, increased cost
and form factor \cite{str1998}, \cite{dfs1}. Multiple input multiple
output (MIMO) and carrier aggregation techniques aggravate the
situation. However, by STR techniques duplexer free systems are
possible by cancelling transmitted signal appeared at receive band.

\subsection{Prior Arts and Proposed Method}

Despite all the advantages mentioned, implementation of STR has its
own challenges. The main difficulty in STR system is echo
cancellation at receive chain. The strongest echo is introduced to
the system when transmitted signal is leaked into receive chain
through circulator. This causes a large interference to the desired
received signal as the echo power level is much higher than desired
received signal. For example, with 46 dBm output power at power
amplifier (PA), assuming 25 dB isolation at circulator, the echo
power will be 21 dBm. In addition to the aforementioned echoes,
echoes can be caused by impedance mismatch at antenna. Besides,
transmitted signal can bounce off objects such as buildings and
mountains, and return to the antenna as echoes at receive chain.
Hence, without echo cancellation, the received signal cannot be
decoded. Furthermore, in order to avoid saturation at LNA and high
resolution ADC, echo cancellation should be performed before LNA.
Therefore, it is imperative to cancel echoes in analogue domain in
order for STR systems to be commercially deployable.

An off-the-shelf hardware \cite{narrow} is available for RF
interference cancellation in analogue and can be used for echo
cancellation in RF. However, it is a narrowband interference
canceller and performs interference cancellation by adjusting phase
and magnitude of interference. Thus, the performance of echo
cancellation in wideband signals is very limited. In \cite{stanford}
about 20 dB echo suppression is reported for 5 MHz IEEE 802.15.4
signals. In addition, this narrowband interference canceller can
create higher sidelobe power than in-band interference power
\cite{stanford}.

An STR implementation in analogue assuming separate transmit and
receive antennas was reported in \cite{str1998} claiming 37 dB
suppression by the RF echo canceller itself. Recently,
\cite{str2012} presented an open-loop technique for full duplex
MIMO. A hardware based experiment shows 50 dB of echo suppression.
In \cite{stanford}, the echo cancellation using transmit beamforming
is studied. This technique uses two transmit antennas and one
receive antenna. The additional two transmit antennas create a null
toward receive antenna, resulting in reduced echo power at receive
chain. However, the transmitted signal is not omni-directionally
transmitted due to the transmit beamforming. When the intended
receiver is located in the null direction, signal-to-noise power
ratio (SNR) loss is inevitable. Additionally, the transmit
beamforming is narrowband beamforming. Thus, it is not suitable for
wideband signal. Since this technique needs separate transmit and
receive antenna set, increased form factor is unavoidable.
Nonetheless, the echo suppression of 20 $\sim$ 30 dB by this
technique is not promising. However, with the aids from digital
baseband cancellation and the narrowband noise canceller
\cite{narrow}, the imperfect prototype achieves 1.84 times of
throughput compared to half duplex system throughput. A scalable
design to MIMO using extra antennas with special placement antennas
is proposed in \cite{midu}. In \cite{mac2}, assuming variable delay
line, one-tap echo canceller is suggested. It is shown that 45 dB
suppression of echo is achieved by using heuristic approach in
adaptive parameter adjustment. The throughput measurements in
\cite{mac2} show 111\% gain in downlink.

Another prior work \cite{rice1} proposed to cancel echo before LNA.
Firstly, to avoid saturation at LNA, separate transmit and receive
antennas with large displacement are used. Various antenna
orientations are investigated in \cite{rice2}. The receiver
estimates the echo channel at every subcarrier using OFDM signal.
Using another transmit radio chain, an echo cancelling OFDM signal
is generated based on the estimated channels of echo path and the
additional radio, and added to the receive chain before LNA. The
measured echo cancellation is about 31 $\sim$ 32 dB in \cite{rice1}
and 24 dB in \cite{rice2}. This type of open-loop technique is
directly sensitive to impairments as it relies on high accurate echo
channel estimation. Since the suppression in this case is not
sufficient to realize the full duplex gain, other mechanisms such as
digital domain cancellation and separate transmit/receive antenna
have to be incorporated.

Without using additional antennas, we propose an adaptive echo
cancellation technique that can be implemented in analogue domain
for wideband signals. The proposed technique is a closed-loop
technique which provides more robustness to impairments than
open-loop techniques. Furthermore, the proposed technique is
scalable to any MIMO system.

\subsection{Applications to Multicell Communications}

As mentioned earlier, with perfect echo cancellation, STR systems
can achieve the doubled capacity especially in isolated links such
as point-to-point communications and wireless backhaul. However, in
cellular systems the situation is different. In addition to the
regular co-channel interference present in half duplex systems,
namely base station (BS) to UE and UE to BS interferences, there are
two unique and serious interferences caused by system operation in
full duplex mode: one is \emph{BS-BS} and the other is \emph{UE-UE}.

On one hand, due to STR at BSs, neighboring BSs' DL signals
interfere with desired UL signal at home BS. This is called
\emph{BS-BS interference} and is extremely severe. Firstly, unlike
BS to UE channel (downlink or uplink), BS-to-BS channel is closer to
line-of-sight (LoS) with much smaller path loss. Secondly, the
transmit power and antenna gain at BS is much larger than those of
UE. Hence, the interferences from neighboring BSs to home BS easily
dominate desired weak UL signal. Hence, without cancelling BS-BS
interferences, UL communication is impossible. In this paper we
suggest a solution for BS-BS interferences and provide system level
evaluations for STR systems with and without BS-BS interference
cancellation.

On the other hand, in STR systems, UL signal transmitted by a UE
creates interference to DL signals received at other UEs
\textit{i.e.}, DL signal will be corrupted by proximate UL signals.
This is called \emph{UE-UE interference} and results in loss of DL
capacity. Unfortunately, with simple transceiver and
omni-directional single antenna at UE, no fancy transmit and receive
beamforming can be utilized in UE to handle UE-UE interference.
Therefore, one option to confront UE-UE interference is to
coordinate between neighboring BSs such that scheduler avoids
scheduling proximate UE pairs that cause serious UE-UE interference
to each other. However, in this paper we propose a non-cooperative
technique which, despite its simplicity, can hugely reduce UE-UE
interference and greatly improve DL capacity.

Unlike cellular systems, the aforesaid BS-BS and UE-UE interferences
are absent in CSMA network such as WLAN systems even when multiple
co-channel cells are deployed. The reason is that due to CSMA
protocol, access points (APs) or terminals will hold their
transmissions if they hear any transmission from other proximate APs
or terminals. In fact owing to STR, the hidden node issue can be
easily solved. Of course collisions still can happen between nodes
due to other factors such as propagation delay and some delay in
header decoding. However, this is common problem even without
employing STR. In this paper we investigate a few protocols and
demonstrate how STR can reduce the hidden node problem. We show that
the throughput is improved even with the assumptions of asynchronous
packet arrival and some delay in decoding header or decision on
channel activity. Unlike \cite{mac1} where detail issues such as
variable packet length, fairness and scheduler design are addressed,
we study the hidden and exposed node issues in simplified model of
CSMA which renders comparisons with theoretical throughput
expression.

The paper is organized as follows: in Section II analogue domain
echo cancellation technique is described together with performance
evaluations. In Section III, the application of STR to CSMA network
for reduction of hidden node problem is discussed and throughput
improvements are demonstrated with multiple cell simulations.
Section IV presents the solutions for BS-BS and UE-UE interferences.
Finally, conclusions are provided in Section V.

\begin{figure*}[!ht]

\centerline{ \subfigure[]
    {\includegraphics[width=3.41 in]{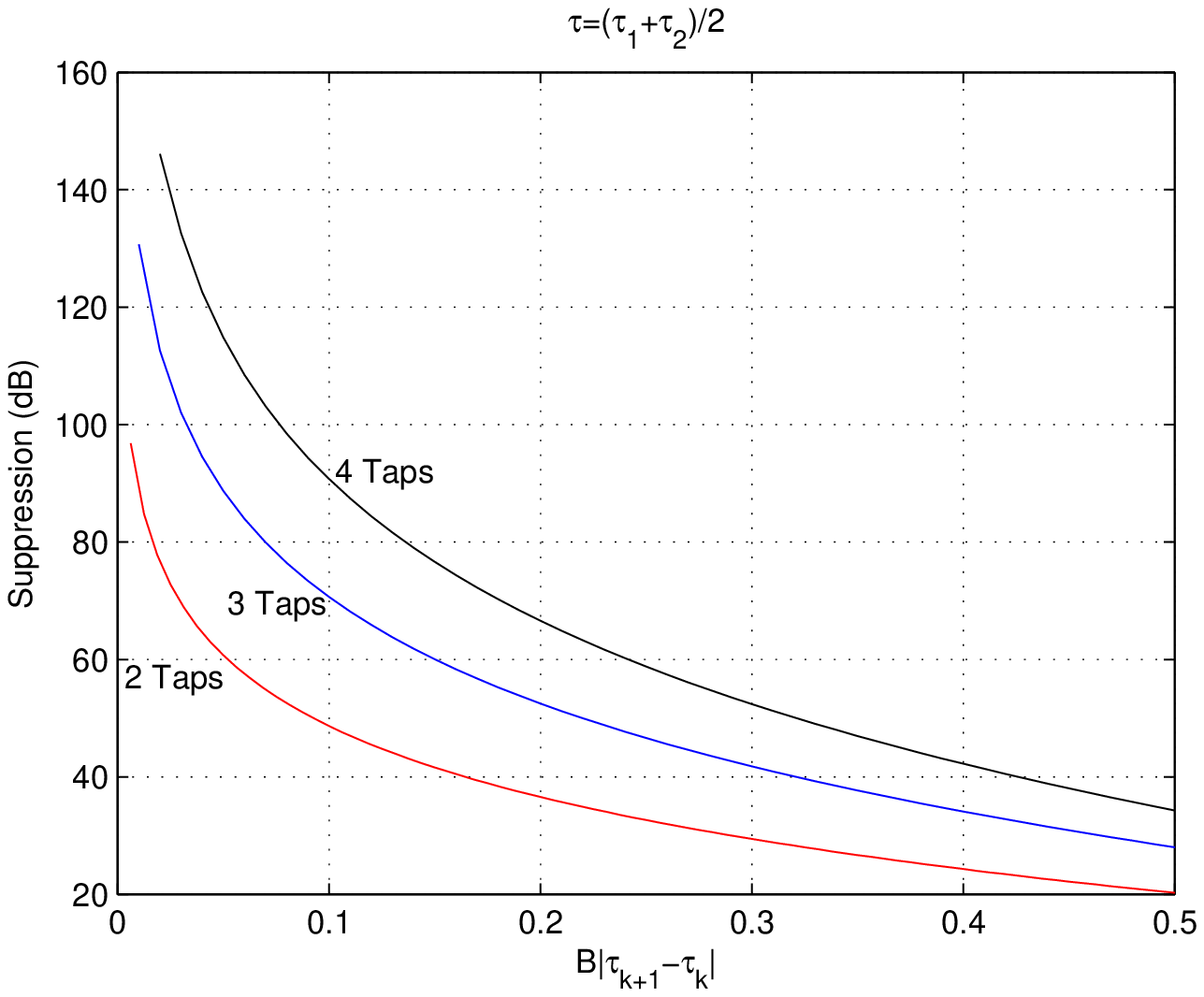}}
    \hfil
\subfigure[]
    {\includegraphics[width=3.41 in]{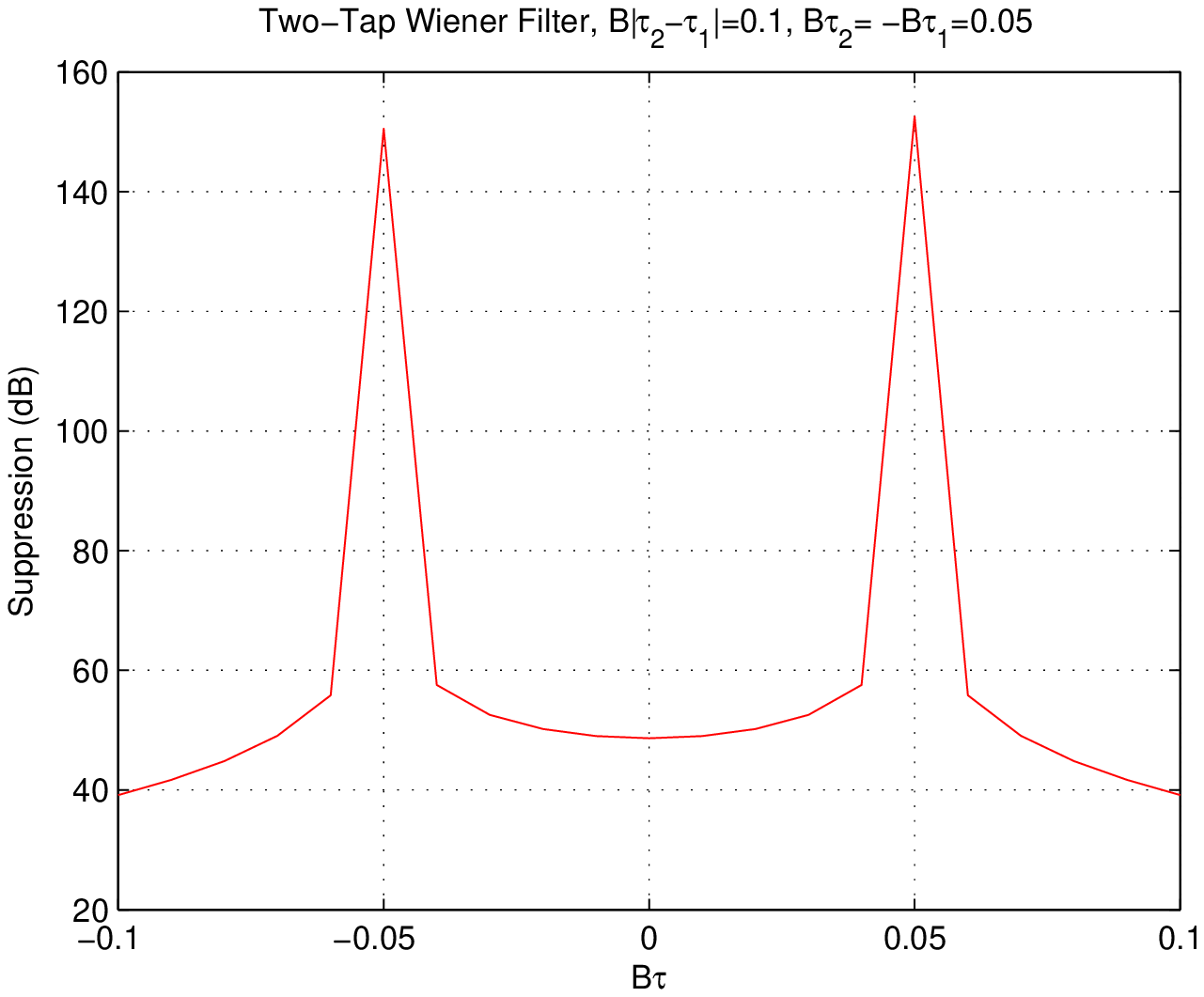}}}

\caption{Performance of Wiener filter: (a) Effect of normalized tap
delay difference and number of taps (b) Effect of echo delay
location in two tap Wiener filter with normalized tap delay
difference equal to 0.1.} \label{fig_mse}

\end{figure*}

\section{Echo Cancellation in Analogue Domain}

Echo cancellation is a well-known topic studied extensively in the
literature and proved technology in the field. Despite the
comprehensive investigations, the derivation of algorithms and
implementations are performed in digital baseband. However, as
described previously in Section I, the echoes should be cancelled
before LNA \emph{i.e.} in analogue domain. Implementing digital
signal processing techniques in analogue domain is very challenging
due to RF impairments and the fact that only limited mathematical
operations are allowed in analogue domain. In this section, we
derive echo cancellation techniques in analogue domain and show that
our method can cancel echoes very effectively even with RF
impairments.

\subsection{System Model and Wiener Solution in Analogue}

The transmitted signal in passband can be written as

\begin{align}
x(t)&= Re\{ \left ( x_i(t) + j x_q(t) \right ) e^{-j\omega t}
\}  \\
    &= x_i(t) \cos (\omega t) + x_q (t) \sin (\omega t)
\end{align}
where $Re\{\cdot\}$ represents the real part, $\omega$ is the
carrier frequency in radian/sec, and $x_i (t)$ and $x_q (t)$ are
baseband in-phase and quadrature phase signals, respectively.
Without loss of generality, we assume one echo with unknown gain and
delay. Then, the received signal $y(t)$ at receive chain can be
modeled as

\begin{equation} \label{y-passband}
y(t)=g x(t-\tau)+r(t)+n(t)
\end{equation}
where $g$ and $\tau$ are the unknown gain and delay of the echo,
respectively, $r(t)$ is the desired received signal and $n(t)$ is
additive white Gaussian noise (AWGN). In this paper we cancel the
echo in passband by creating estimated echo signal using multiple
replica of $x(t)$ with different delays and subtracting it from the
signal corrupted by echo. For instance, the echo can be estimated by
a linear combination of $x(t-\tau_1)$ and $x(t-\tau_2)$ where
$\tau_k$ is the delay of $k$-th tap. In order to rotate the phases
of $x(t-\tau_1)$ and $x(t-\tau_2)$, we also need to use the Hilbert
transform of $x(t-\tau_1)$ and $x(t-\tau_2)$. Then, the estimated
echo using two taps can be written as

\begin{equation}\label{e-t0}
e(t)=w_{1i} x(t-\tau_1)+w_{1q} \hat{x} (t-\tau_1)+ w_{2i}
x(t-\tau_2)+w_{2q} \hat{x} (t-\tau_2)
\end{equation}
where $w_{ki}$ and $w_{kq}$ are in-phase and quadrature component of
$k$-th tap weight, and $\hat{x}(t)=x_i (t) \sin (\omega t) - x_q (t)
\cos (\omega t)$ is the Hilbert transform of $x(t)$. Then, the echo
canceller output can be expressed as

\begin{equation} \label{cancellOutput0}
z(t)=y(t) - e(t).
\end{equation}

In order to simplify derivation of the weights $w_{ki}$ and
$w_{kq}$, we transform the passband model to equivalent complex
baseband model. The transmitted complex baseband signal $X(t)$ is
defined by

\begin{equation}
X(t)=x_i (t) + j x_q (t).
\end{equation}
Then, from \eqref{y-passband}, the received signal in complex
baseband is in the form of

\begin{equation}
Y(t)=g X (t-\tau)e^{j\omega \tau} + R(t)+N(t)
\end{equation}
where $R(t)$ and $N(t)$ are complex baseband versions of desired
received signal and AWGN, respectively. Likewise, we can develop the
complex baseband model of the echo canceller output from
(\ref{e-t0}) and (\ref{cancellOutput0}) as follows

\begin{equation}
Z(t)=Y(t)-W_1 X (t-\tau_1)e^{j\omega \tau_1} -W_2 X
(t-\tau_2)e^{j\omega \tau_2}.
\end{equation}
After some manipulations, it is not difficult to show that
$W_k=w_{ki}+j w_{kq}$. For the concise representation, the echo
canceller output can be written in vector form as

\begin{equation}
Z(t)=Y(t)-\textbf{\textit{W}}^T \textbf{\textit{X}} (t)
\end{equation}
where superscript $(\cdot)^T$ represents transpose,
$\textbf{\textit{W}}= [ W_1~ W_2]^T$ and $\textbf{\textit{X}} (t)=[
X (t-\tau_1)e^{j\omega \tau_1}~ X (t-\tau_2)e^{j\omega \tau_2} ]^T$.

Although it is not possible to implement Wiener solution in analogue
domain, minimum mean squared error (MMSE) solution provides valuable
insights on the performance and design parameters. Minimizing the
power of the echo canceller output means that the echo is cancelled.
This is due to the fact that unknown desired receive signal $r(t)$
and AWGN cannot be cancelled by linear combination of
$\{x(t-\tau_k)\}$ as $\{x(t-\tau_k)\}$ is uncorrelated with $r(t)$
and noise. Therefore, we define the cost function which minimizes
the power of the echo canceller output as

\begin{equation} \label{cost}
\min_{\textbf{\textit{W}}} E\{ | Z(t) |^2
\}=\min_{\textbf{\textit{W}}} E\{ | Y(t)-\textbf{\textit{W}}^T
\textbf{\textit{X}} (t) |^2 \}.
\end{equation}
In Appendix A the derivation of Wiener solution in analogue domain
is presented. Echo cancellation performance is shown in Fig.
\ref{fig_mse} where the suppression level is defined by

\begin{align}
\text{Suppression} &= \frac{\text{Echo Power}}{\text{Residual Echo
Power}} \label{sup}\\
&=\frac{E\{ |g X(t-\tau) |^2  \}}{E \{ |g X(t-\tau) e^{j\omega \tau}
- \textbf{\textit{W}}^T \textbf{\textit{X}} (t) |^2 \}}.\label{res}
\end{align}

\noindent The suppression is not a function of the carrier frequency
and echo power as shown in Appendix A. Rather, it is a function of
delay difference between taps and the number of them. Fig. 1 (a)
exhibits suppression as a function of tap delay difference
normalized by signal bandwidth (B) assuming that the echo delay is
in the middle of the first and second tap delay. As it can be
observed from the figure, smaller normalized tap delay difference
$\left (B|\tau_2-\tau_1| \right )$ and larger number of taps provide
better echo suppression. Even with two taps, the suppression level
close to 90 dB can be achieved with the normalized tap delay
difference equal to 0.01 which means 1 nsec of delay difference
between taps in 10 MHz bandwidth systems. This is equivalent to 100
times oversampling in digital signal processing. It is apparent that
as the tap delay difference goes to zero the suppression level goes
to infinity (\emph{i.e.} no residual echo). We emphasize that having
a small tap delay difference is not a challenging issue. For
example, assuming the speed of electromagnetic (EM) wave in
transmission line is equal to the speed of light, $3mm$ difference
in length will yield $10$ pico second delay difference. In Fig.
\ref{fig_mse} (b) the effect of echo delay on echo canceller
performance is illustrated assuming that the normalized tap delay
difference is 0.1. In reality the delay of echo is unknown. However,
typical value or range of values of circulator delay can be
measured. Interestingly, when the tap delay coincides with the true
delay of the echo, even with only two taps suppression beyond 150 dB
can be easily achieved. The reason that even when $\tau$ is equal to
either $\tau_1$ or $\tau_2$ the suppression is not infinity is that
$\tau_2-\tau_1$ is not approaching zero. It is simple to show that
when the delay of either taps coincides with the echo delay, one tap
echo canceller instead of two tap echo canceller has infinity
suppression level.

In two tap Wiener filter, we study the impact of number of echoes
under the condition of same total echo powers. Assume one echo
having delay $\tau$ located in the middle of neighboring two taps

\begin{equation}
Y_1(t)= X (t-\tau)e^{j\omega \tau}
\end{equation}
where $\tau=(\tau_1+\tau_2)/2$. Without loss of generality assume
$E\{|X(t)|^2 \}=1$. Let us add one more echo with the same power as
the first echo but closer to the second tap with arbitrary phase
$\vartheta$ and delay $\tau+\delta$ where $0 < \delta<
(\tau_2-\tau_1)/2$ and $\tau_2>\tau_1$. Then, the received echo
signal becomes

\begin{equation}
Y_2(t) =  X (t-\tau)e^{j\omega \tau} + X (t-\tau-\delta)e^{j\omega
(\tau+\delta)+\vartheta}.
\end{equation}
The power of $Y_2(t)$ is

\begin{equation}
E\{ |Y_2(t)|^2 \}= 2 \left ( 1 +  \cos
(\omega\delta+\vartheta)Sinc(B\delta) \right ).
\end{equation}
Depending on the phase $\omega\delta+\vartheta$, the new echo can
reduce total echo power. However, we consider the worst case
\emph{i.e.} $\cos (\omega\delta+\vartheta)=1$. Note
$Sinc(B\delta)>0$ due to the assumption $B\delta<1$. For fair
comparison, we normalize $Y_2(t)$ such that maximum power of
$Y_2(t)$ is equal to the power of $Y_1(t)$

\begin{equation}
Y_2(t) = c \left ( X (t-\tau)e^{j\omega \tau} + X
(t-\tau-\delta)e^{j\omega \tau} \right )
\end{equation}
where $c=1/\sqrt{2+2Sinc(B\delta)}$. Then, $Y_1(t)$ and $Y_2(t)$
have the same power but $Y_1(t)$ has one echo in the middle of taps
while $Y_2(t)$ contains two echoes: one in the middle of the taps
and the other closer to either taps.

The cross correlation vector can be written as

\begin{equation}
\textbf{\textit{R}}_{Y_2\textbf{\textit{X}}}=E\{ Y_2(t)
\textbf{\textit{X}}^H(t)\} = c e^{j\omega \tau}
(\textbf{\textit{r}}_{{}_{Y\textbf{\textit{X}}}} +
\textbf{\textit{r}}_{{}_{Y_2\textbf{\textit{X}}}})\textbf{\textit{D}}
\end{equation}
where $\textbf{\textit{r}}_{{}_{Y\textbf{\textit{X}}}}$ and
$\textbf{\textit{D}}$ are defined in Appendix A and
\begin{equation}
\textbf{\textit{r}}_{{}_{Y_2\textbf{\textit{X}}}}=[Sinc(B(\tau+\delta-\tau_1))~Sinc(B(\tau+\delta-\tau_2))].
\end{equation}

It is not difficult to show the residual echo powers of $Y_1(t)$ and
$Y_2(t)$ as following

\begin{eqnarray}\label{resEcho1}
mse_1&{}={}&1-
\textbf{\textit{r}}_{{}_{Y\textbf{\textit{X}}}}\textbf{\textit{r}}_{{}_{\textbf{\textit{XX}}}}^{-1}\textbf{\textit{r}}_{{}_{Y\textbf{\textit{X}}}}^H
\end{eqnarray}
and,
\begin{equation}\label{resEcho2}
mse_2=\frac{mse_1+mse_0+2(Sinc(B\delta)-\textbf{\textit{r}}_{{}_{Y_2\textbf{\textit{X}}}}\textbf{\textit{r}}_{{}_{\textbf{\textit{XX}}}}^{-1}\textbf{\textit{r}}_{{}_{Y\textbf{\textit{X}}}}^H)}{2+2Sinc(B\delta)},
\end{equation}
respectively where $mse_0=1-
\textbf{\textit{r}}_{{}_{Y_2\textbf{\textit{X}}}}\textbf{\textit{r}}_{{}_{\textbf{\textit{XX}}}}^{-1}\textbf{\textit{r}}_{{}_{Y_2\textbf{\textit{X}}}}^H$
which is the residual echo power of the second echo when the first
echo is not present.

\begin{figure}[!t]

\centering
    {\includegraphics[width=3.41 in]{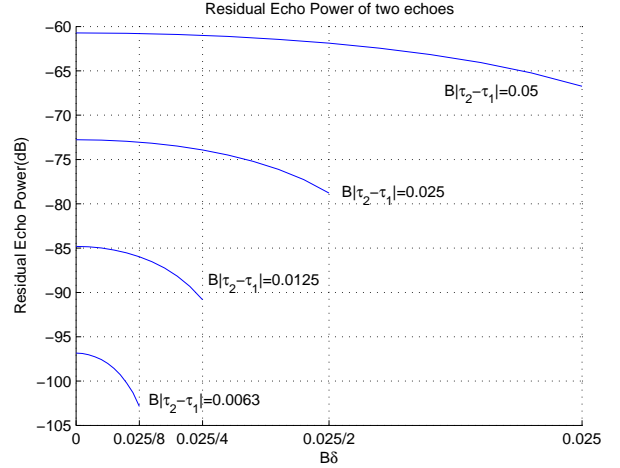}}

\caption{Effect of two echoes in two tap Wiener filter.
\label{fig_mse2echoes} }

\end{figure}

From Fig. \ref{fig_mse2echoes} it is clear that as the second echo
moves toward the second tap, the residual echo power is reduced.
Notice that $\delta=0$ means that only one echo exists. Hence, the
assumption of one echo in the middle of neighboring taps is the
worst simulation condition. This is from the fact that echo closer
to either tap is easier to estimate as shown in Fig. \ref{fig_mse}
(b). The argument can be easily extended to arbitrary number of
echoes with arbitrary echo powers located between neighboring taps.

\begin{figure}[!ht]

\centering \subfigure[] {
    {\includegraphics[width=3.41 in]{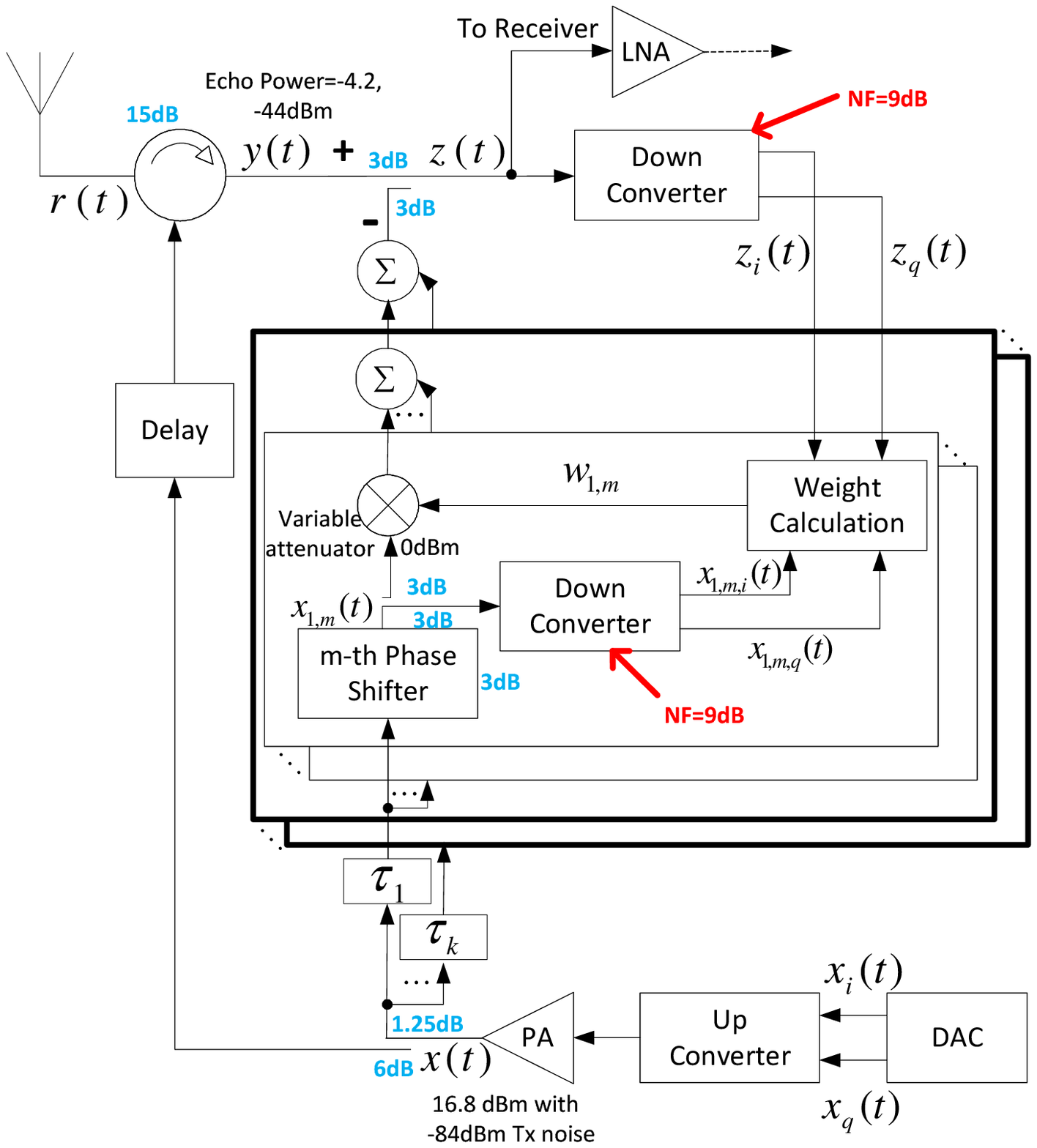}} }
\subfigure[] {
    {\includegraphics[width=2.8 in]{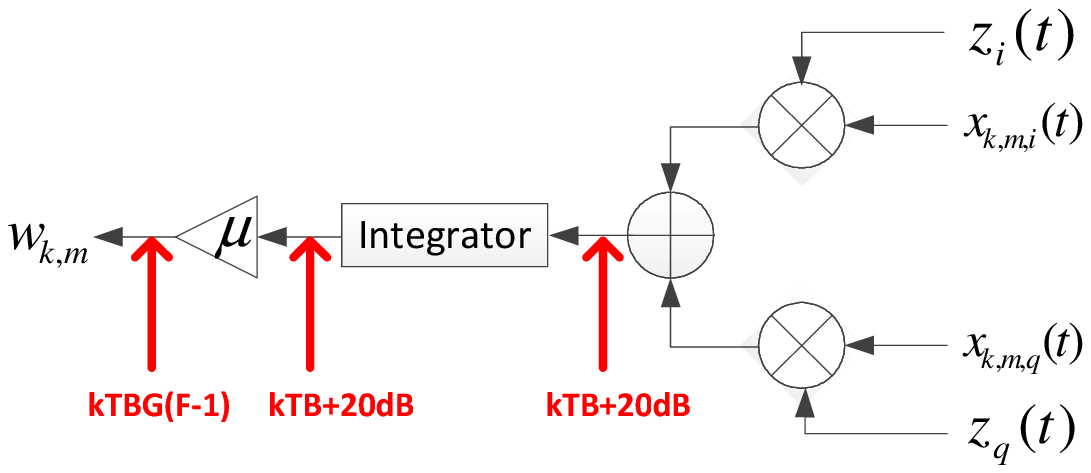}} }

\caption{Block diagram of adaptive echo cancellation in analogue
domain: (a) High level block diagram (b) Weight calculation block.
\label{fig_blk} }

\end{figure}

\subsection{Adaptive Echo Cancellation in Analogue}

In the previous subsection ideal phase shifter is assumed for
Hilbert transform. However, due to phase imbalance in phase shifter,
Hilbert transform of $x(t-\tau_k)$ will not be orthogonal to
$x(t-\tau_k)$. To overcome the phase imbalance problem we use
multiple phase shifters. As far as those phases are not identical,
by a linear combination of the multiple phase shifter outputs we can
rotate and scale $x(t-\tau_k)$ even with large phase imbalances.
Assuming $K$ taps and $M$ phase shifters per tap, $m$-th phase
shifter output at $k$-th tap can be written as

\begin{align}\label{x-t}
x_{k,m}(t)= g_{k,m} \{ & x_i(t-\tau_k) \cos (\omega (t-\tau_k)
-\theta_{k,m}) \nonumber \\
+& x_q(t-\tau_k) \sin (\omega (t-\tau_k) -\theta_{k,m})\}
\end{align}
where $g_{k,m}$ is phase shifter gain to represent amplitude
imbalance, and $\theta_{k,m}$, which can be modeled as $\pi
m/M+\phi_{k,m}$, is phase shift including phase imbalance
$\phi_{k,m}$. It is clear that with $M=3$, as far as the phase
imbalance is less than $30^\circ$, those three phases will not be
identical. By a linear combination of phase shifter outputs, echo
can be estimated as follows

\begin{equation}\label{e-t}
e(t)= \sum_{k=1}^{K} \sum_{m=1}^{M} w_{k,m} x_{k,m}(t)
\end{equation}
where $w_{k,m}$ is real valued weight for $k$-th tap and $m$-th
phase shifter. The complex baseband model of echo canceller output
$z(t)$ can be developed from (\ref{cancellOutput0}) and (\ref{e-t})
as

\begin{equation}
Z(t)=Y(t)-\sum_{k=1}^{K} \sum_{m=1}^{M} w_{k,m} X_{k,m}(t).
\end{equation}
where $X_{k,m}(t)$ is complex baseband version of $x_{k,m}(t)$. The
real and imaginary parts of $X_{k,m}(t)$ can be written as

\begin{align}\label{X-kmi}
X_{k,m,i}(t)= g_{k,m} \{ & x_i(t-\tau_k) \cos ( \omega \tau_k
+\theta_{k,m}) \nonumber \\
-& x_q(t-\tau_k) \sin (\omega \tau_k +\theta_{k,m})\},
\end{align}

and
\begin{align}\label{X-kmq}
X_{k,m,q}(t)= g_{k,m} \{ & x_i(t-\tau_k) \sin ( \omega \tau_k
+\theta_{k,m}) \nonumber \\
+& x_q(t-\tau_k) \cos (\omega \tau_k +\theta_{k,m})\},
\end{align}
respectively. Since implementing Wiener filter in analogue domain is
quite difficult if not impossible, it is desirable to design
techniques which are implementable in analogue domain. For this end,
we apply well-known steepest-descent method to the cost function
\eqref{cost}. This leads to

\begin{align}
w_{k,m}&= w_{k,m}-\frac{\mu}{2} \frac{\partial | Z(t) |^2} {\partial
w_{k,m}} \nonumber \\
&=w_{k,m}+ \mu Re\{ X_{k,m}^*(t) Z(t) \}\label{lms}
\end{align}

\noindent where $\mu$ is step-size and superscript $(\cdot)^*$
represents complex conjugate. Notice that cross correlation $Re\{
X_{k,m}^*(t) Z(t) \}$ can be served as echo channel estimation at
$k$-th tap delay of $m$-th phase shifter output. Since it is quite
noisy estimation due to random signal $x(t)$, noise and impairments
which will be discussed in next subsection, the noisy estimation is
low-pass filtered by integrator with small step-size $\mu$ as shown
in (\ref{lms}).

Together with \eqref{lms}, \eqref{cancellOutput0} leads to the block
diagram of the adaptive echo cancellation in analogue domain shown
in Fig. \ref{fig_blk}. Note that the multipliers in the outputs of
phase shifters are in fact variable attenuators since the magnitudes
of weights can be limited to less than 1 by design. Also notice that
as the step-size $\mu$ is much less than 1, it can be implemented by
a fixed attenuator. Overall, it is quite clear that the proposed
structure is implementable.

\subsection{Impairments}

Unfortunately, however, $\{X_{k,m}(t)\}$ and $Z(t)$ cannot be
obtained due to phase noise, phase/amplitude imbalances in
downconverter. With unknown phase error $\phi_{z,i}$ and
$\phi_{z,q}$ in in-phase and quadrature phase respectively, and
unknown gains of $g_{z,i}$ and $g_{z,q}$ in in-phase and quadrature
phase respectively, following signal has to be used

\begin{align}
\widetilde{Z}(t)= g_{z,i}&Re\{ Z(t) e^{-j
(\phi_{z,i}+\varphi_z(t))}\} \nonumber \\
+ jg_{z,q}&Im\{ Z(t) e^{-j (\phi_{z,q} +\varphi_z(t)) }\}
\end{align}
where $Im\{\cdot\}$ represents
imaginary part and $\varphi_z(t)$ is phase noise. Similarly, we can
define $\widetilde{X}_{k,m}(t)$. In addition, although all
downconverters are driven by the same oscillator, each downconverter
may have different phase rotation. Hence, effectively following will
be performed

\begin{align}
w_{k,m}&=w_{k,m}+ \mu Re\{ \widetilde{X}_{k,m}^*(t) \widetilde{Z}(t)
e^{j\Delta_{k,m}}\} \label{Delta}
\end{align}
where $\Delta_{k,m}$ is the phase difference between downconverter
for $Z(t)$ and downconverter for $X_{k,m}(t)$. It is desirable to
have $\Delta_{k,m}<45^\circ$ in order to avoid divergence or slow
convergence. Next, the effect of phase distortion in the variable
attenuator is investigated. In wideband variable attenuator
\cite{rfmd}, different attenuation causes different phase rotation
$e^{j\phi(w_{k,m})}$. Thus the second term in the right-hand side of
\eqref{Delta} is changed to

\begin{align}
\mu Re\left \{ \widetilde{X}_{k,m}^*(t) \widetilde{Z}(t)
e^{j(\Delta_{k,m}-\phi(w_{k,m}))}\left (1-jw_{k,m}
\frac{\phi(w_{k,m})} {\partial w_{k,m} } \right) \right\}.
\end{align}
As in \cite{rfmd}, the phase distortion can be modeled as a linear
function of attenuation in log scale. For example, assuming
$0^\circ$ and $50^\circ$ phase rotations at 3 dB and 38 dB
attenuations respectively leads to

\begin{align}\label{weightC}
w_{k,m}&=w_{k,m}+ \mu Re\left \{ \widetilde{X}_{k,m}^*(t)
\widetilde{Z}(t) e^{j\left (\Delta_{k,m}-\phi(w_{k,m})+12^\circ
\right)} \right\}.
\end{align}
The additional phase rotation due to the phase distortion in the
attenuator can be absorbed in $\Delta_{k,m}$. The phase distortion
is large when the attenuation is large. However, when the
attenuation is large, its contribution to echo estimation is small.
Thus, the phase distortion at high attenuation can be ignored. The
values of above mentioned impairments are unknown. Hence, we do not
take any attempt to compensate those impairments in our study.
Simply, the outputs of downconverters which include all those
impairments are used to train the weights.

Although the second terms in the right hand side of \eqref{lms} and
\eqref{weightC} are quite different due to impairments, as far as
the signs of the second terms are same, the adaptive technique can
converge as explained in Appendix B. This is why the proposed method
exhibits immunity to impairments unlike open loop techniques.

The fixed phase shifter, combiner, splitter and coupler are quite
linear components. In \cite{rfmd}, the input third order intercept
point (IIP3) of the variable attenuator is 50 dBm which means with 0
dBm input power non-linear distortion power can be less than -100
dBm. Hence, 0 dBm input power at the variable attenuator is
maintained.

\begin{center}
\begin{table*}[ht]
\centering \caption{Phase/Amplitude imbalances : In the order of
Phase shifter/Downconverter I/Q}
\begin{small}
\begin{tabular}{c||c|c|c||c|c|c}
  \hline
  \hline
&\multicolumn{3}{c||}{Phase Imbalances \& $\Delta_{k,m}$}  &\multicolumn{3}{c}{Amplitude Imbalances}  \\
$k$ &\multicolumn{3}{c||}{$\phi_{k,m}$, $\phi_{x,k,m,i}$, $\phi_{x,k,m,q}$, $\Delta_{k,m}$}  &\multicolumn{3}{c}{$g_{k,m}$, $g_{x,k,m,i}$, $g_{x,k,m,q}$}  \\
  \cline{2-7}
  & $m=1$ & $m=2$ & $m=3$ & $m=1$ & $m=2$ & $m=3$   \\
  \hline
  \hline
  Tap 1 & -1, 0.6, 0.3, 30$^\circ$ & 10, 0.1, 1.1, -30$^\circ$ & 5, 0.25, -0.75, 15$^\circ$ & -0.5, 0.2, 0.7 dB  & 1, 0.1, 0.6 dB & 1, 0.23, -0.27 dB   \\
  Tap 2 & 11, -0.5, 0.5, 10$^\circ$ & -2, 0.5, -0.4, 23$^\circ$ & 2, -0.15, 0.8, -17$^\circ$ & -1, -0.1, 0.4 dB  & 0.5, 0.5, -0.1 dB & 2, 1, 0.5 dB   \\
  Tap 3 & 2, -0.8, 0.2, -25$^\circ$ & 12, -0.2, 0.8, 19$^\circ$ & -5, 0.7, -0.2, -30$^\circ$ & 1.5, -0.3, 0.2 dB  &-0.2, -0.2, 0.3 dB & 2, 0.4, -0.1 dB   \\
  \hline
  \hline
\end{tabular}
\end{small}
\end{table*}
\end{center}
At the output of every component, noise with power of $kTBG(F-1)$ is
added where $kTB$ represents thermal noise power over bandwidth $B$,
$G$ is power gain and $F$ is noise factor. In passive components
including the variable attenuator $F=1/G$ is assumed. The loss of
power in component is specified in Fig. \ref{fig_blk}. $N$-way
combiner and splitter are assumed to have loss of $10\log_{10} N$
per branch (not shown in Fig. \ref{fig_blk}). The noise figure of
downconverter is assumed to be 9 dB. As shown in Fig. \ref{fig_blk}
(b), 20 dB of noise above thermal noise is added at the integrator
output. For the noise in baseband multipliers and adder we
equivalently add 20 dB of noise above thermal noise.

BS and AP may not need to support multiple bands in a given operator
and deployment. However, terminal will need to support multiple
bands. Multiband support is more pronounced in cellular systems. It
is more challenging to have good antenna match over multiple bands
in terminal. Hence, dominant echo can come from antenna mismatching
in terminals. Considering terminal, we assume echo power of only 15
dB lower than transmit power mainly due to antenna mismatching
although typical broadband commercial circulator has isolation of 20
$\sim$ 25 dB with fractional dB of insertion loss. Notice that BS
and AP can accommodate bulky circulator which can provide more than
40 dB isolation and will need to tune antenna for fewer number of
bands usually. By better antenna match and smaller leakage in
circulator, the 6 and 3 dB loss of signal power in couplers in
transmit and receive path respectively can be reduced by increasing
the loss of coupler in echo estimation path.

Table I shows phase imbalances in phase shifters ($\phi_{k,m}$) and
corresponding downconverters' in-phase and quadrature phase
imbalances($\phi_{x,k,m,i},\phi_{x,k,m,q}$) relative to in-phase of
transmit signal. The phase rotation difference $\Delta_{k,m}$ is
specified as well. Amplitude imbalances in phase shifters
($g_{k,m}$) and corresponding downconverters' in-phase and
quadrature phase amplitude imbalances ($g_{x,k,m,i},g_{x,k,m,q}$)
relative to in-phase of transmit signal are tabulated in Table I. In
the downconverter for $z(t)$, phase imbalances of $-0.9^\circ$ and
$0.1^\circ$, and amplitude imbalances of -0.1 dB and 0.4 dB are
assumed for in-phase and quadrature phase, respectively. In transmit
chain, it is assumed that the quadrature phase has $1.5^\circ$ phase
and 0.8 dB amplitude imbalance relative to in-phase. Note that phase
and amplitude imbalances larger than commercially available
components are assumed. Regarding DC-offset, zero DC-offset is
considered which can be achieved by DC blocking capacitor.

From the derivation of adaptive echo cancellation, no restriction on
transmitted signal is imposed. Hence, even if transmitted signal
$x(t)$ is contaminated by impairments such as phase/amplitude
imbalance, non-linear distortion, phase noise, and additional noise
above thermal noise, the performance in analogue domain is
unaffected as long as the bandwidth is remained the same. Of course,
when the signal bandwidth is changed due to impairments, the tap
delay difference $|\tau_{k+1}-\tau_k|$ should be adjusted
accordingly to maintain the same normalized tap delay difference. In
adaptive echo cancellation and Wiener filter, the echo $x(t-\tau)$
is estimated by linear combination of $\{x(t-\tau_k)\}$. Thus, it is
important for the input signal to echo estimation path to be
identical to the input signal to echo channel path. However, in
digital domain echo cancellation, this property cannot be met. This
is a significant advantage over digital domain echo cancellation
where the digital baseband signal which goes to echo estimation path
does not contain all RF impairments in transmit chain.

\subsection{Simulations}

Instead of training sequence, 10 MHz random OFDM signals with 102
active subcarriers and QPSK symbols are transmitted. A white noise
20 dB above thermal noise is added to OFDM signal right after power
amplifier as a transmit noise. At the output of circulator two
echoes are created in simulations and an echo canceller with three
taps is considered. The first echo with -4.2 dBm power is located in
the middle of the first and second tap delay, and the second echo
with -44 dBm power is located in the middle of the second and third
tap delay. We set the normalized tap delay difference $B
|\tau_{k+1}-\tau_k|$ to be 0.025 which is equivalent of 40 times
oversampling in delay domain. This means that the delay spread is
2.5 nsec. Note that with three taps we can cancel echoes of 5 nsec
delay spread at least.

\begin{figure*}[!ht]

\centerline{ \subfigure[]
    {\includegraphics[width=3.41 in]{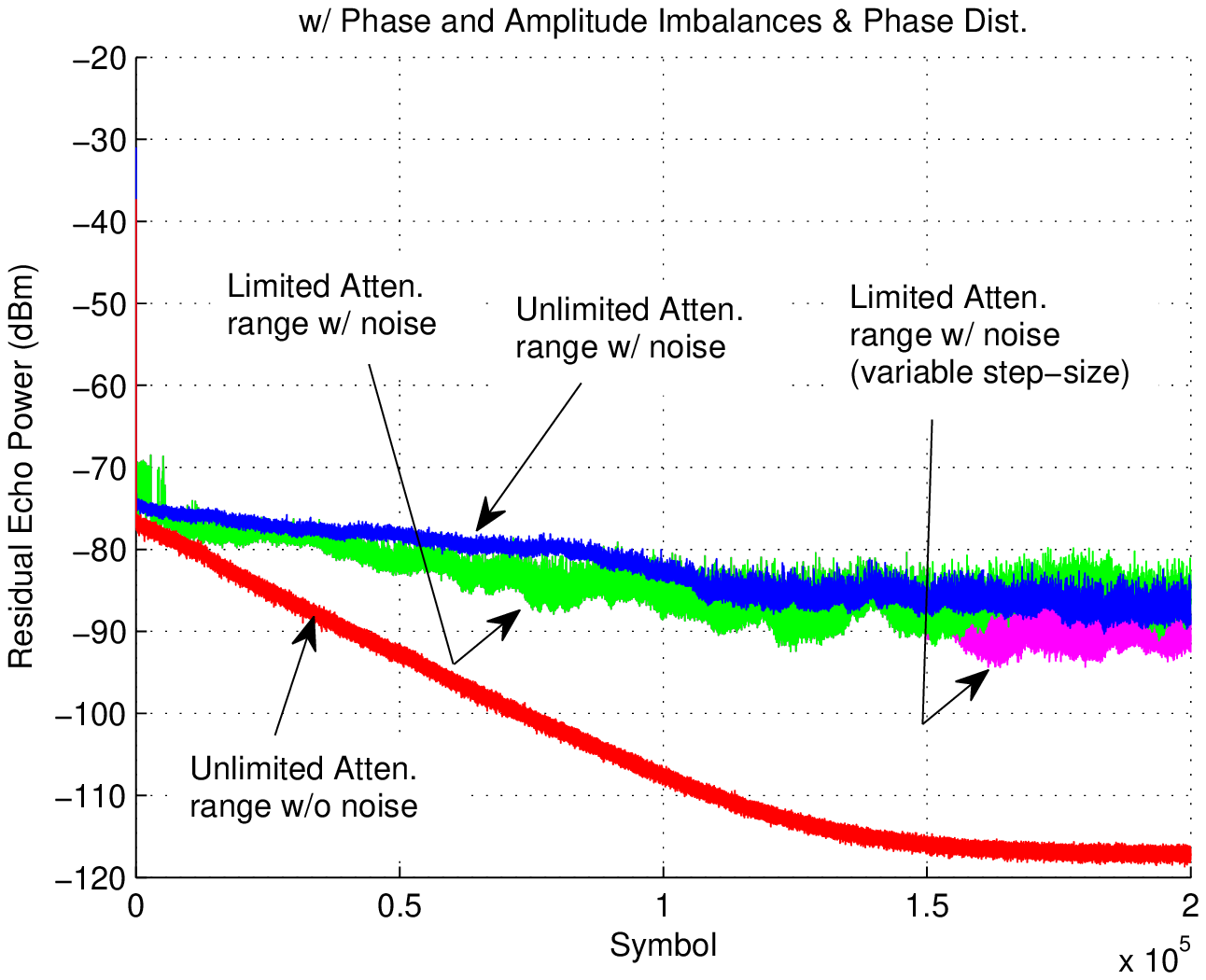}}
    \hfil
\subfigure[]
    {\includegraphics[width=3.41 in]{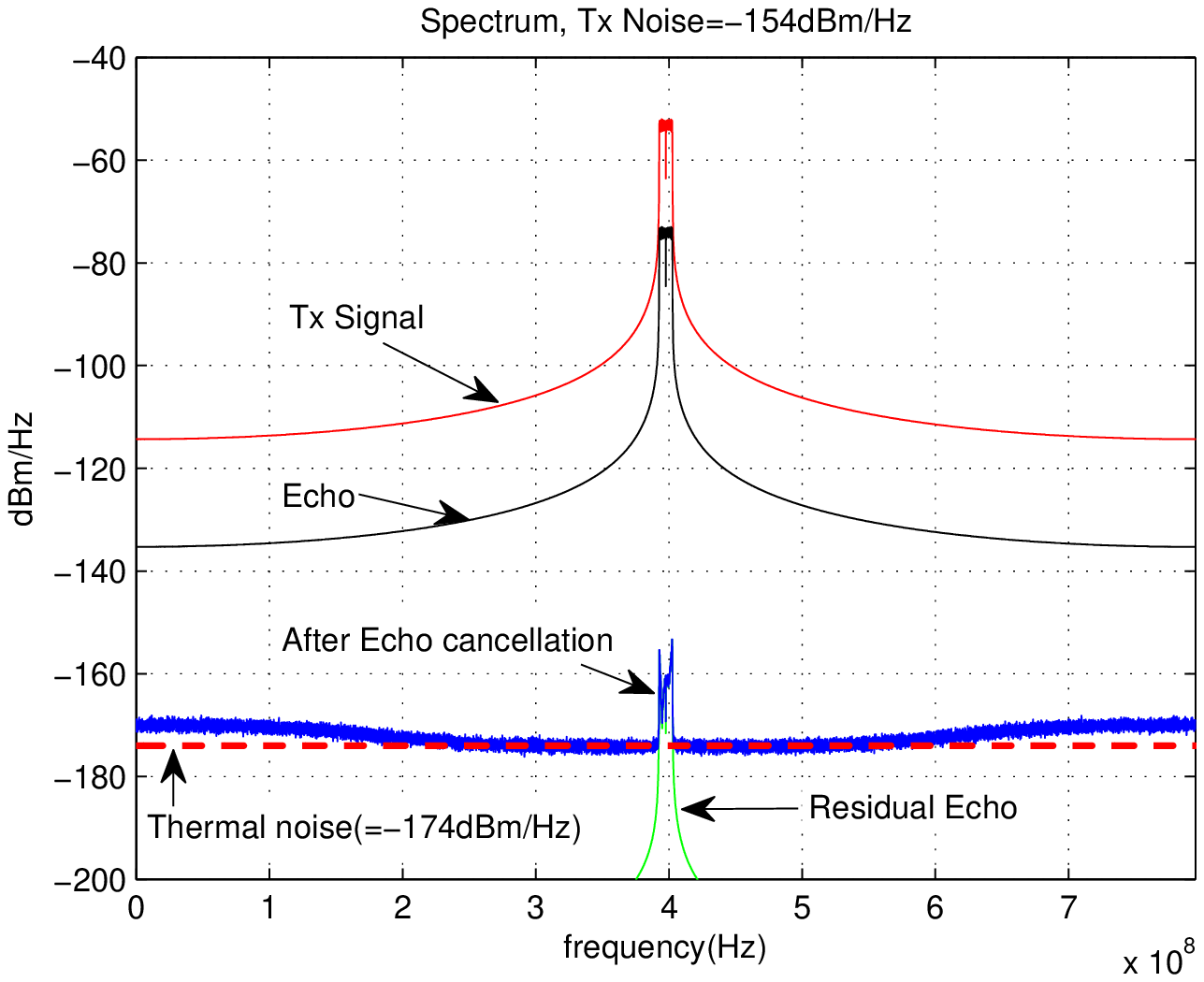}}}

\caption{Adaptive analogue echo canceller with 3 taps,
$B|\tau_{k+1}-\tau_k |=0.025$ and two echoes (-4.2 and -44 dBm): (a)
Convergence (b) Spectrum.} \label{fig_conv}

\end{figure*}

In Fig. \ref{fig_conv} (a) simulation results of adaptive echo
cancellation in analogue domain are demonstrated. Assuming 12.5\% of
guard interval $2\times 10^5$ symbols translate 2.32 sec.
Aforementioned all impairments excluding phase noise are assumed.
When unlimited dynamic range is assumed in the variable attenuator
without noise, the in-band residual echo power is far less than
thermal noise power -104 dBm. In this case, more than 110 dB of
suppression which is close to Wiener filter solution is possible
(see Fig. \ref{fig_mse} (a)). When impairments do not limit the
performance, the adaptive echo cancellation can converge to Wiener
solution. The suppression level in Wiener filter is not a function
of absolute power of echo but the number of taps and the normalized
tap delay difference. The result demonstrates that the
phase/amplitude imbalances and phase distortion are not limiting
factors.

When RF impairments do not limit performance, it is not difficult to
show that more suppression of echo can be achieved by smaller
normalized tap delay difference as demonstrated in Wiener filter
section. This is one of benefits in analogue domain echo
cancellation since high oversampling for digital domain processing
is not feasible. However, by decreasing the normalized tap delay
difference, convergence becomes slower while echo cancellation
performance improves. This slower convergence is typical behavior of
larger eigenvalue spread of auto-correlation matrix of tap input in
adaptive filter \cite{haykin}. It is clear that by decreasing the
normalized tap delay difference, the eigenvalue spread of the
auto-correlation matrix of tap input \eqref{auto} becomes larger.

By adding noise, however, we observe slow convergence and higher
floor. When the dynamic range of the attenuator is limited to 35
dB(\emph{i.e.} 3 $\sim$ 38 dB attenuation), fluctuation is
noticeable since the limited dynamic range will interfere the
convergence behavior especially when weights have to change signs.
It is clear that as the residual echo power becomes smaller, the
variance of the residual echo power increases. This is due to the
fact that as the echo power is reduced, SNR from echo point of view
worsens. When noise becomes effective, the variance can be reduced
by smaller step-size \cite{haykin}. As seen in Fig. \ref{fig_conv}
(a) the residual echo power can be less than -90 dBm when the
step-size is reduced by half at $1.5\times 10^5$ symbol. Simple
noise analysis shows that heavy noise at weight calculation block
(20 dB of noise before and after integrator) governs the absolute
value of residual echo power as it reduces the SNR of residual echo
and directly affects the weight calculation (\ref{weightC}). It is
critical to reduce the noise at baseband and improve the dynamic
range of the attenuation. Fig. \ref{fig_conv} (b) exhibits the
spectrum when the residual echo power is -90 dBm. It is interesting
to note that the out-of-band power of $z(t)$ at vicinity of in-band
is equal to thermal noise (=-174 dBm/Hz) even if transmit noise is
-154 dBm/Hz. This demonstrates that the transmit noise above thermal
noise can be cancelled as well. However, the noise power at
frequencies far from in-band is a bit higher than thermal noise.
Note that the cost function \eqref{cost} will be driven by high
power component in frequency. Hence, the weights are mainly
influenced by in-band signal. Thus, as far as out-of-band power is
smaller than in-band power, weights will be chosen to reduce in-band
power. In addition, the normalized delay difference is not
sufficiently small to cancel out-of-band noise. Therefore, less
reduction of echo in out-of-band is possible. It is clear that the
proposed structure is enough for duplexer free systems and device
co-existence problem. However, the residual echo power is too high
to operate STR unless the noise at baseband is decreased and the
dynamic range in the attenuator is improved.

\begin{figure}[!t]

\centering
    {\includegraphics[width=3.41 in]{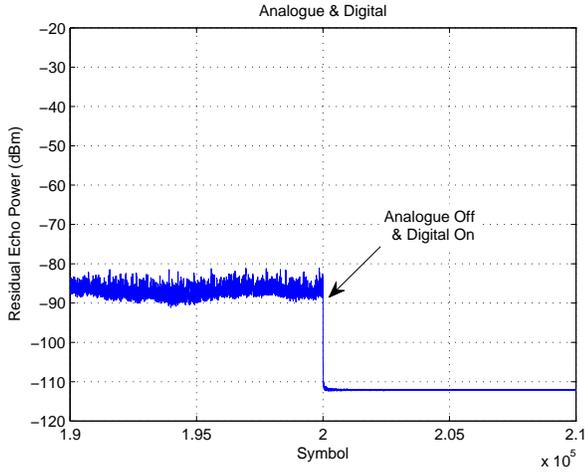}}

\caption{Analogue and digital echo cancellation. \label{fig_digital}
}

\end{figure}

We bring attention to the prototypes in \cite{stanford},
\cite{mac2}, \cite{rice1} where additional 20 $\sim$ 30 dB of echo
suppression is achieved by digital domain echo cancellation. When
the power of $z(t)$ is less than a threshold, the analogue weight
update is disconnected and digital domain echo cancellation is
enabled. By disconnecting from analogue update circuit, the heavy
noise from baseband and other impairments will not appear in receive
chain via weights. Also by turning off the analogue weight
calculation block and multiple downconverters, power consumption can
be reduced. In this paper simple open-loop technique is applied for
digital domain echo cancellation. The echo channel at every
subcarrier is estimated and this estimation is averaged over time.
From the estimated echo channel, echo signal is estimated at every
subcarrier and subtracted from received signal. The joint simulation
of analogue and digital echo cancellation is shown in Fig.
\ref{fig_digital}. It is clear that the residual echo power is less
than thermal noise. In fact due to large noise figure in receive
chain (5 dB in BS and 9 dB in UE for example), the background noise
will be -99 dBm and -95 dBm respectively. This will give us more
tolerance in phase/amplitude imbalance in transmit chain.

\subsection{Phase noise and Time varying channel}

\begin{figure}[!t]

\centering
    {\includegraphics[width=3.41 in]{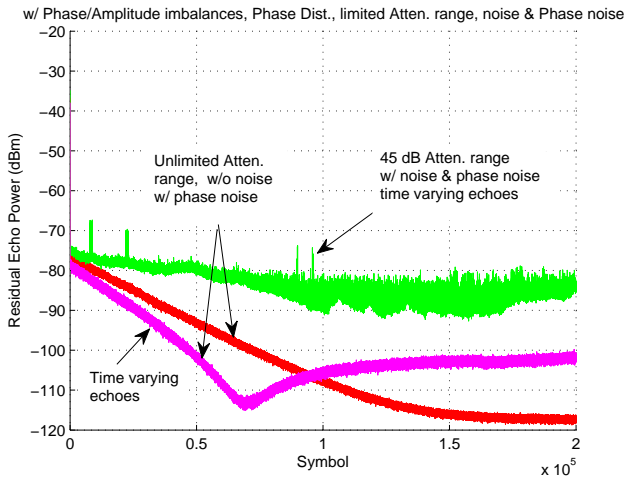}}

\caption{Convergence with white phase noise ($\sigma=2^\circ$) and
time varying echoes. \label{fig_convPhaseNoise} }

\end{figure}

In this subsection, phase noise is added on top of the previous
simulation conditions. Phase noise is assumed to be white Gaussian
noise with root mean square (rms) equal to 2 degree. All
downconverters are assumed to have independent phase noises. From
Fig. \ref{fig_convPhaseNoise}, when noise free and unlimited
attenuation range in variable attenuator are assumed, the
convergence behavior is almost identical to the case without phase
noise (see Fig. \ref{fig_conv} (a)). It is evident that the white
phase noise is not a limiting factor. As explained in Appendix B,
the phase noise effectively reduces the average power of echo
channel estimation (\ref{EErrWPN}). This means that the SNR from
residual echo point of view is smaller when noise is present. Hence,
in this case noise at baseband becomes more prominent than the case
without phase noise. The main reason why the proposed method is less
sensitive to phase noise is that the phase noise does not directly
appear at weights. Rather, the phase noise is low-pass filtered with
small step-size $\mu$ which controls the bandwidth of the low pass
filter. Hence, unlike open-loop techniques, the closed-loop
techniques such as adaptive filter has good level of immunity to
impairments. In \cite{phasenoise} an analysis on the phase noise
impact to echo suppression is reported when open-loop technique is
applied. From the analysis, only 26 dB of echo suppression is
possible when rms phase noise is $2^\circ$. Note that open-loop
technique is directly sensitive to various impairments as
impairments deteriorate the accuracy of echo channel estimation.

\begin{figure}[!t]

\centering
    {\includegraphics[width=3.41 in]{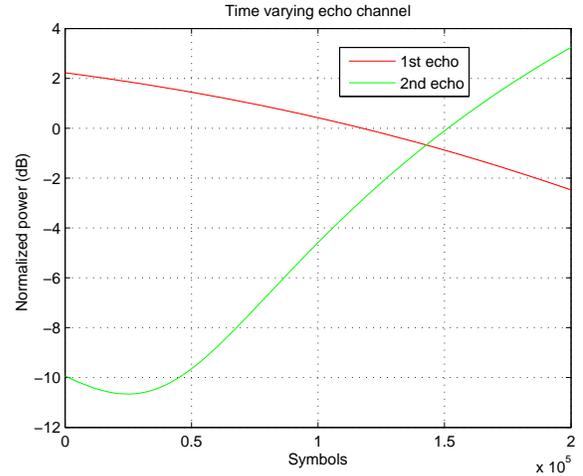}}

\caption{Time varying echo channel. \label{fig_tvh} }

\end{figure}

Next, we consider the effect of time varying channel. The echoes
reflected from objects in time varying wireless channel and coming
back to receive antenna have large path loss due to round trip.
Since these objects introduce additional loss and also orientations
of the objects are not always positioned toward receive antenna, in
general these echoes have relatively small powers. In all other
proposals which employ separate transmit and receive antenna, all
echoes, including the strongest echo, come from over-the-air
channel. Therefore, the strongest echo can be more time variant
compared to our proposal. This is due to the fact that, in our
proposal the strongest echo comes from RF components inside device
which usually has good shielding and consequently, much less time
variation of echo channel is expected. Nonetheless, to demonstrate
the capability of tracking time varying echoes, we simulate time
varying echo channels. The time varying impulse response is shown in
Fig. \ref{fig_tvh}. The first echo has 4.7 dB of dynamic range while
the second echo has 14 dB of power change over $2\times 10^5$ OFDM
symbols. First, let us focus on the case when unlimited attenuation
range and noise free are assumed. Since the prediction of time
varying channel is not possible, higher residual echo power is
unavoidable. In addition, the second echo power is growing. Thus,
when the residual power of the second echo becomes more dominant
than that of the first echo, the residual echo power is increased as
well since the supersession level is fixed regardless of absolute
power of echo. For faster tracking the step-size is increased in
this simulation.

As addressed earlier limited range of attenuation interferes smooth
change of sign of weights. Hence, glitches in convergence are
observed as seen in Fig. \ref{fig_conv} (a). Although adaptive
algorithm quickly stabilizes the convergence owing to negative
feedback, for smoother tracking of time varying channel it is
desirable to increase the attenuation range. Therefore, we assumed
45 dB range of attenuation. As seen in Fig.
\ref{fig_convPhaseNoise}, even in this severe time varying channel,
our proposed technique converges very well. In general, adaptive
filter has good level of capability in tracking time varying channel
owing to closed-loop operation. In open-loop technique, the echo
cancellation performance is directly coupled with echo channel
estimation error. For better echo cancellation, more accurate
channel estimation is required and consequently, we will need to
average the impulse response for longer period of time. This means
that the capability of tracking time varying channel is
fundamentally limited. However, in closed-loop technique, when the
residual echo signal $Z(t)=Y(t)-E(t)$ becomes suddenly large due to
any reason, the power of second term in right hand side of
(\ref{lms}) which drives the weights is increased as well.
Therefore, the weights are immediately changed to reduce the power
$|Z(t)|^2$.

\subsection{Discussions}

Although in the simulations two echoes are assumed, multiple echoes
with same delay can be treated as one echo. In reality the delay
spread of echoes should be measured, and first and second tap delays
should be decided in order to cover the delay spread. If the tap
delay difference is too large, more taps are needed between these
two taps in order to make sure the normalized tap delay difference
between neighboring taps $B|\tau_{k+1}-\tau_k|$ to be sufficiently
small. For any pair of neighboring taps, we added only one echo with
delay equal to the middle of two neighboring taps. As demonstrated
earlier, this is the worst case scenario. The residual echo power is
less dependent on the number of echoes but more on the total echo
power and normalized delay difference $B|\tau-\tau_k|$ where $\tau$
is echo delay and $\tau_k$ is the delay of nearest tap. In order to
have sufficient degrees of freedom, the number of taps should be
large enough to cover the delay spread and minimize normalized delay
difference $B|\tau-\tau_k|$.

We emphasize that RF implementations for adaptive duplexers or
duplexer free systems \cite{dfs1}, \cite{dfs2} are similar to our
proposed structure except weight calculation algorithm as those use
the same building blocks such as phase shifters and variable
attenuators in passband. In addition, echoes appear at receive chain
via either circulator or duplexer. However, those prototypes make a
null at a particular frequency, which means that the algorithm does
not take care of other frequencies. Thus, it cannot create wide null
and out-of-band power can be large. In contrast, our scheme chooses
weights which minimize the power of echo canceller output over whole
band. Hence, out-of-band residual echo power is not higher than
in-band residual echo power. Of course, by filtering $z(t)$ weights
can be optimized for a particular band. It is interesting to note
RFIC product \cite{motia}, \cite{motia2} in which adaptive blind
receive beamforming is implemented in analogue domain. From
\cite{motia2} it is evident that weights are adaptively calculated
in analogue baseband and applied in passband to achieve receive
beamforming in RF. The high level structure which implements
adaptive algorithm in \cite{motia2} is similar to our proposal
although the objective and detail algorithms are different.

From simulation results within 20 $\sim$ 30 symbols (0.23 $\sim$
0.35 msec) the residual echo powers reach -70 $\sim$ -77 dBm. Note
that the weights are initialized to lowest positive values which are
irrelevant to echo channel impulse response. However, the weights
are quickly trained to reduce the residual echo power to -70 $\sim$
-77 dBm. For extra 10 $\sim$ 20 dB suppression, 1 $\sim$ 2 $\times
10^5$ symbols are needed. For faster convergence, variable step-size
can be applied. Initially, the algorithm can start with large
step-size (meaning large bandwidth of low pass filter) for faster
initial convergence and later the step-size can be reduced for
refinement and reducing noise impact. By well-designed training
sequence, fast convergence and small variation of residual echo
power can be obtained as well. Adaptation of weights can be done in
initial start-up of the device or during production. Moreover, lower
power training sequence can be used to update the weights before
entering network. Note that unlike echo channel in an echo canceller
that employs separate transmit and receive antennas, echo channel at
circulator can vary over time with much slower rates. Therefore, it
is sufficient to calculate weights in the initial power up. In order
to compensate any possible drift of circulator channel, it is
adequate to update the weights during preamble or for a short period
during which no reception of signal is expected or the received
power is small. Since updating weights does not have to be performed
always, additional power consumption due to multiple downconverters
and weight calculation blocks can be insignificant.

We demonstrated the feasibility of the echo canceller implementation
in analogue domain with RF impairments. Given degrees of freedom
such as number of taps and tap delay difference, the adaptive
technique searches best weights such that echo power is minimized
over entire band. Impairments will interfere convergence and may
increase residual echo power. However, the negative feedback
mechanism provides a certain extent of immunity to impairments. When
baseband noise is high and dynamic range of variable attenuator is
limited, digital baseband echo cancellation can be accompanied. It
is also clear that the proposed technique can be easily extended to
any MIMO system. However, one should note that in that case the echo
cancellation complexity will be exponentially increased due to the
cross talks coming from adjacent transmit antennas. Nonetheless,
applying analogue echo cancellation to MIMO systems is not
impossible. It is apparent from the formulations that the echo does
not have to be the same type of radio signal as received signal. For
example $r(t)$ can be WLAN signal while $x(t)$ can be Bluetooth
signal.

\section{MAC Level Simulations in CSMA Networks}

\begin{figure}[!t]

\centering \subfigure[] {
    {\includegraphics[scale=1.1]{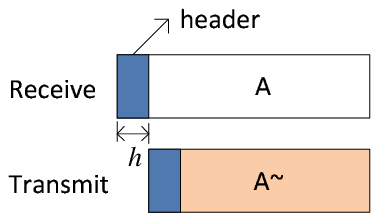}} }
\subfigure[] {
    {\includegraphics[scale=1.1]{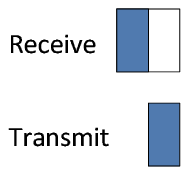}} }
\subfigure[]{
    {\includegraphics[scale=1.1]{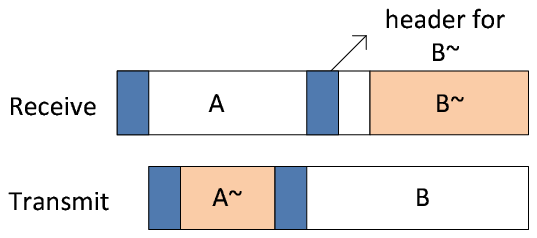}} }

\caption{Frame (a) s-STR: send a dummy packet(A$\sim$) upon
successful reception of the header (b) i-STR: example of collision
(c) d-STR: asynchronous arrival of transmit and receive packets.
\label{frame} }

\end{figure}

In this section the benefits of STR in CSMA networks are studied. We
explore both single cell and multiple cells in order to study the
effect of STR in reducing the hidden node issue and improving
throughput. It is assumed that any node has the same coverage radius
$r$ and signal is not reached beyond the radius. Hence, any two
nodes within a distance of $r$ can sense each other's channel
activity. The hidden node problem is the major source of collisions.
In a cell, downlink signal can be detected over entire cell
coverage. Therefore, during downlink no collision will happen.
However, in uplink when one terminal is located at cell edge while
other terminal is located at the cell edge in the opposite side of
the cell, these two terminals are hidden from each other. Thus, the
packets from these two nodes can collide with each other. It is hard
to eliminate the hidden node issue in CSMA networks due to the
nature of channel sensing mechanism. However, it is clear that STR
can reduce the hidden node issue. In fact, in a single cell the
hidden node problem can be completely resolved by employing STR and
sending a dummy packet. More precisely, as illustrated in Fig.
\ref{frame} (a), whenever an AP or terminal receives a designated
packet (A), if it sends a dummy packet (A$\sim$), hidden nodes can
sense the channel as busy and therefore they will hold packets. We
refer to this technique as \textit{s-STR}.

Unfortunately, there is another source of collision. In real world
systems, it takes some time for any node to detect energy, or decode
a header and make a decision on channel activity. In other words, it
is not instant to decide whether the channel is busy or idle. Let
$h$ be the elapsed time for detecting the energy of received signal
or decoding a header represented in percentage of packet duration.
When $h$ is non-zero, some collisions occur that are unavoidable
even by means of STR. Of course in a single cell scenario with STR
capability and only one active terminal, no collision occurs even
with non-zero $h$. However, in general the collisions due to
non-zero $h$ overshadows the gain of STR.

In order to further improve collision avoidance, we introduce
\textit{i-STR} which utilizes the dummy packet as an implicit ACK.
Unlike traditional ACK, the implicit ACK is very immediate with the
delay of only $h$ while traditional ACK has the delay of at least
one packet duration. We can devise a protocol based on receiving the
implicit ACK or not. For example, a destination node receiving a
packet will send a dummy packet (\textit{i.e.} implicit ACK) only
during the period of the arriving packet if the destination
successfully decodes the header of the packet. If the source node
can detect the header of the implicit ACK from the destination node,
it continues its current transmission. Otherwise, it stops the
current transmission. Likewise, when the destination node cannot
decode the header, it will not transmit the implicit ACK. Then, the
source node stops the transmission. This modification will release
the channel much earlier than simply sending the packet and
corresponding dummy packet over the whole duration of the packet. As
a result, other potential nodes can access the channel. After
successful reception of the header at the destination node, as in
s-STR, a dummy packet is sent by the destination node. Fig.
\ref{frame} (b) shows an example of i-STR when the header of dummy
packet was collided at source node. Then, the source cannot detect
the header of the implicit ACK. It decides no ACK transmission from
the destination at time $2h$. The source stops packet transmission
at time $2h$. Then, the destination node stops dummy packet
transmission as well. By this fast interaction and stopping
transmission when collision is detected, instead of wasting the
channel, other potential nodes can utilize it.

\begin{figure*}[!t]

\centerline{ \subfigure[]
    {\includegraphics[width=3.0 in]{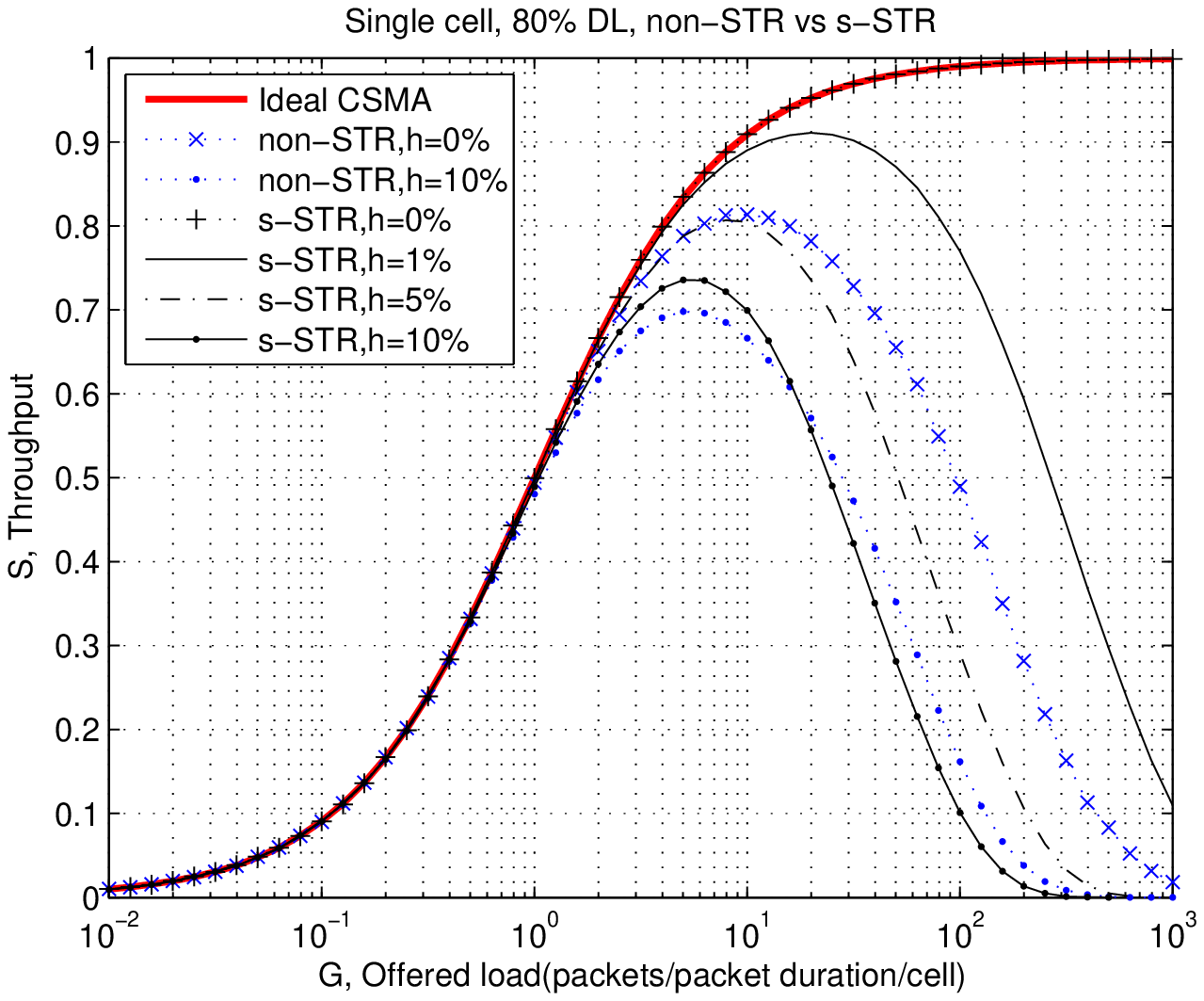}}
\hfil \subfigure[]
    {\includegraphics[width=3.0 in]{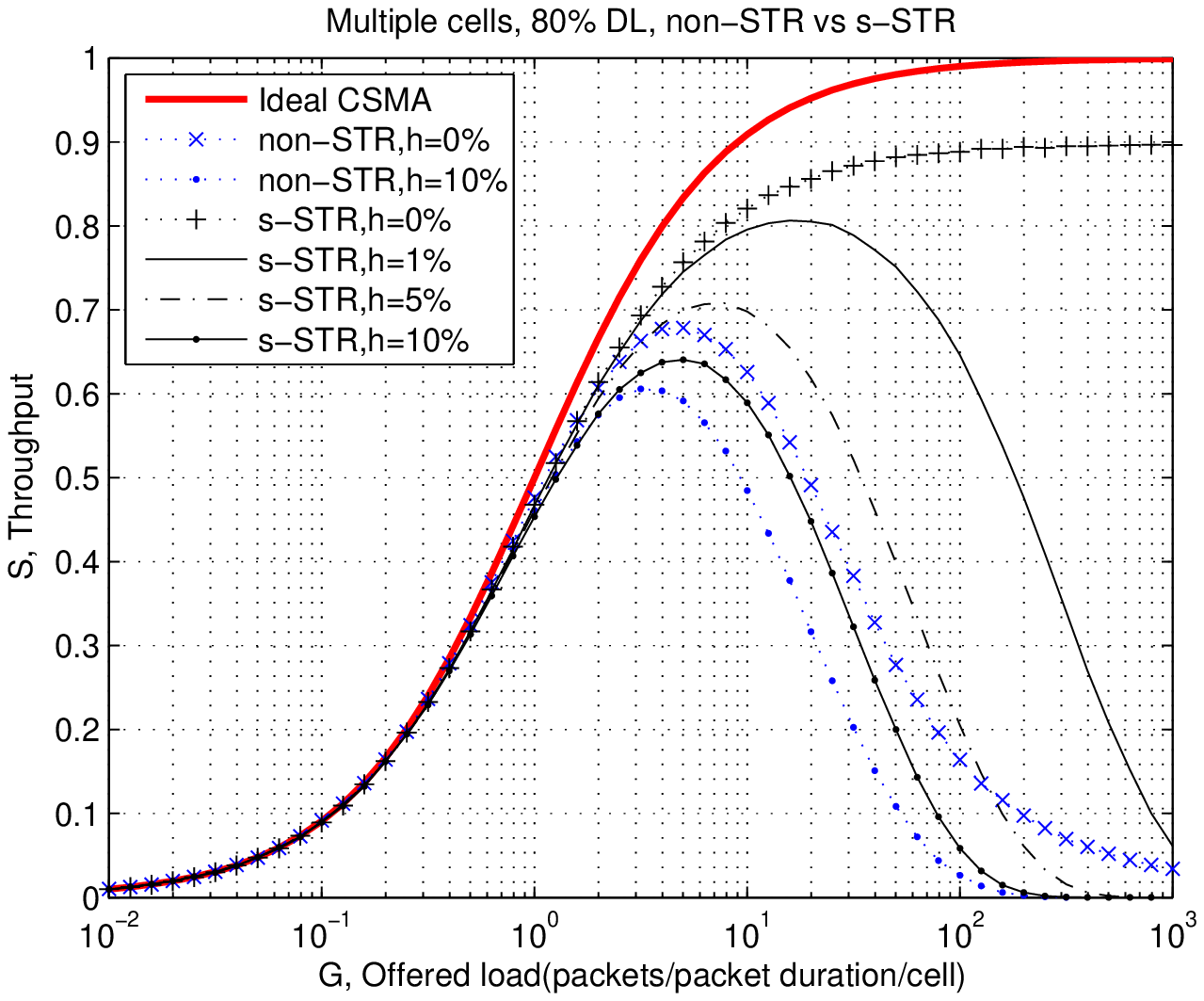}}
} \centerline{ \subfigure[]
    {\includegraphics[width=3.0 in]{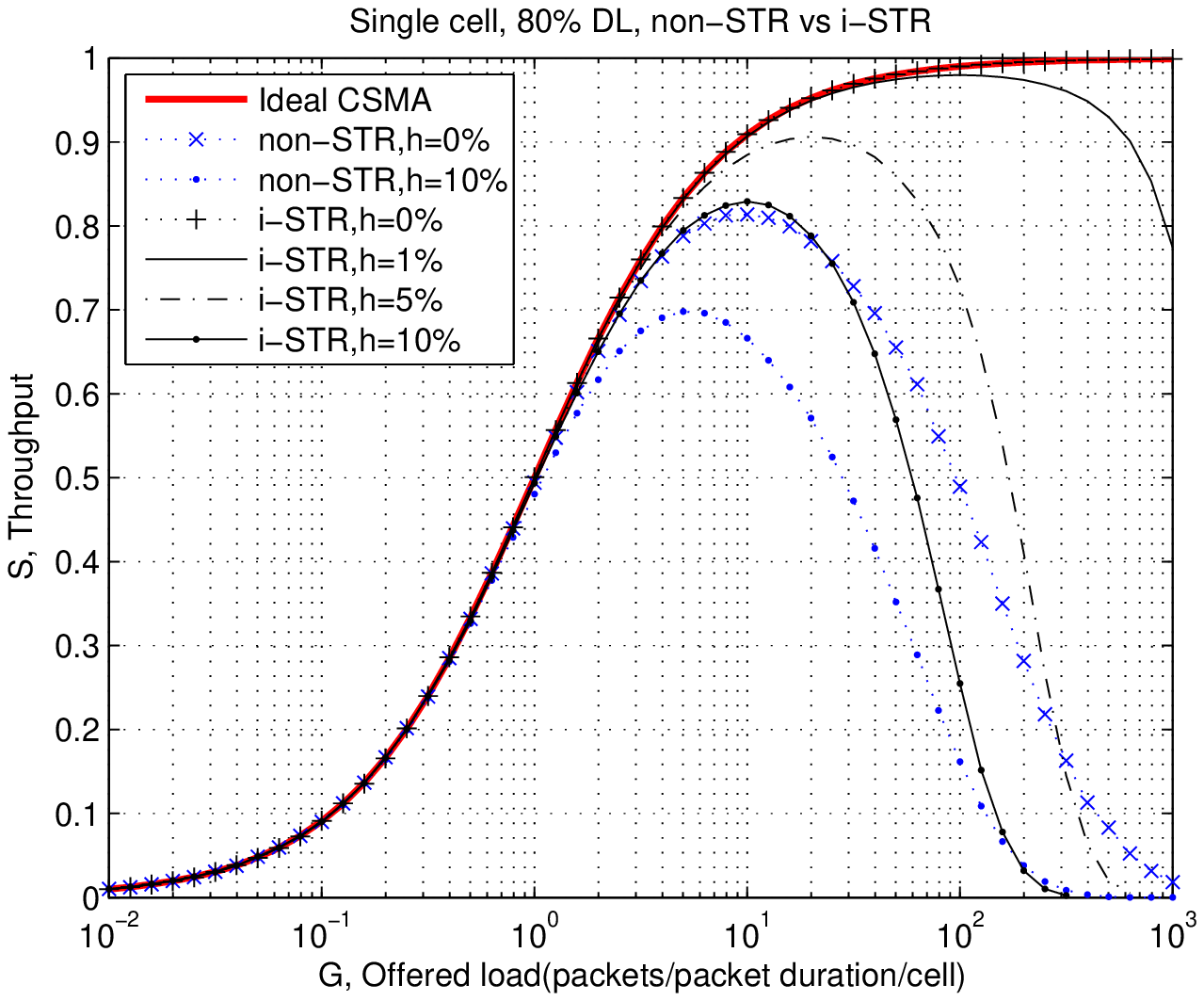}}
\hfil \subfigure[]
    {\includegraphics[width=3.0 in]{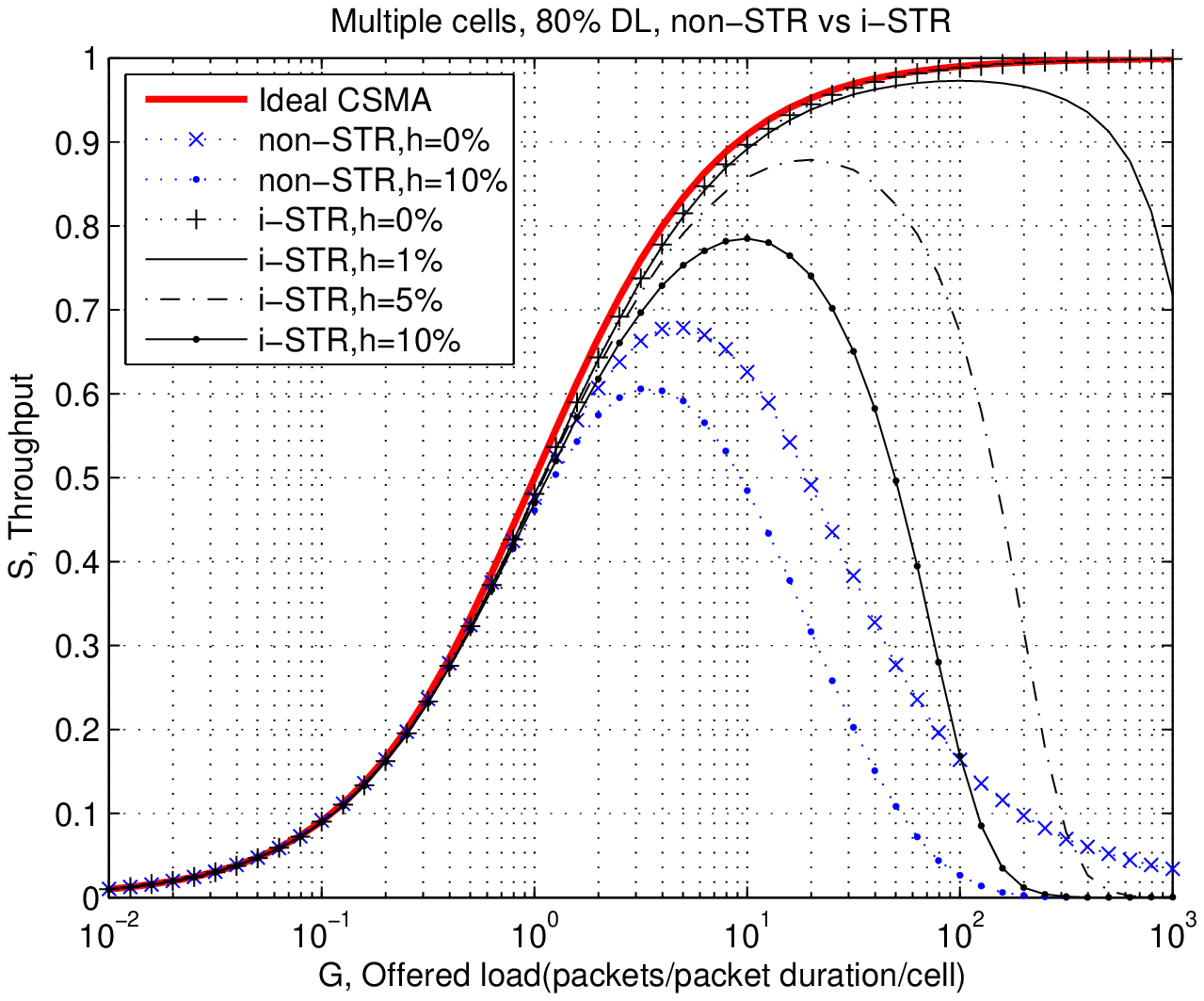}}}

\caption{Throughput comparisons, DL traffic=80\%: (a) s-STR in
single cell (b) s-STR in multiple cells (c) i-STR in single cell (d)
i-STR in multiple cells. } \label{fig_s_i_STR}

\end{figure*}

\begin{figure*}[!ht]

\centerline{ \subfigure[]
    {\includegraphics[width=3.0 in]{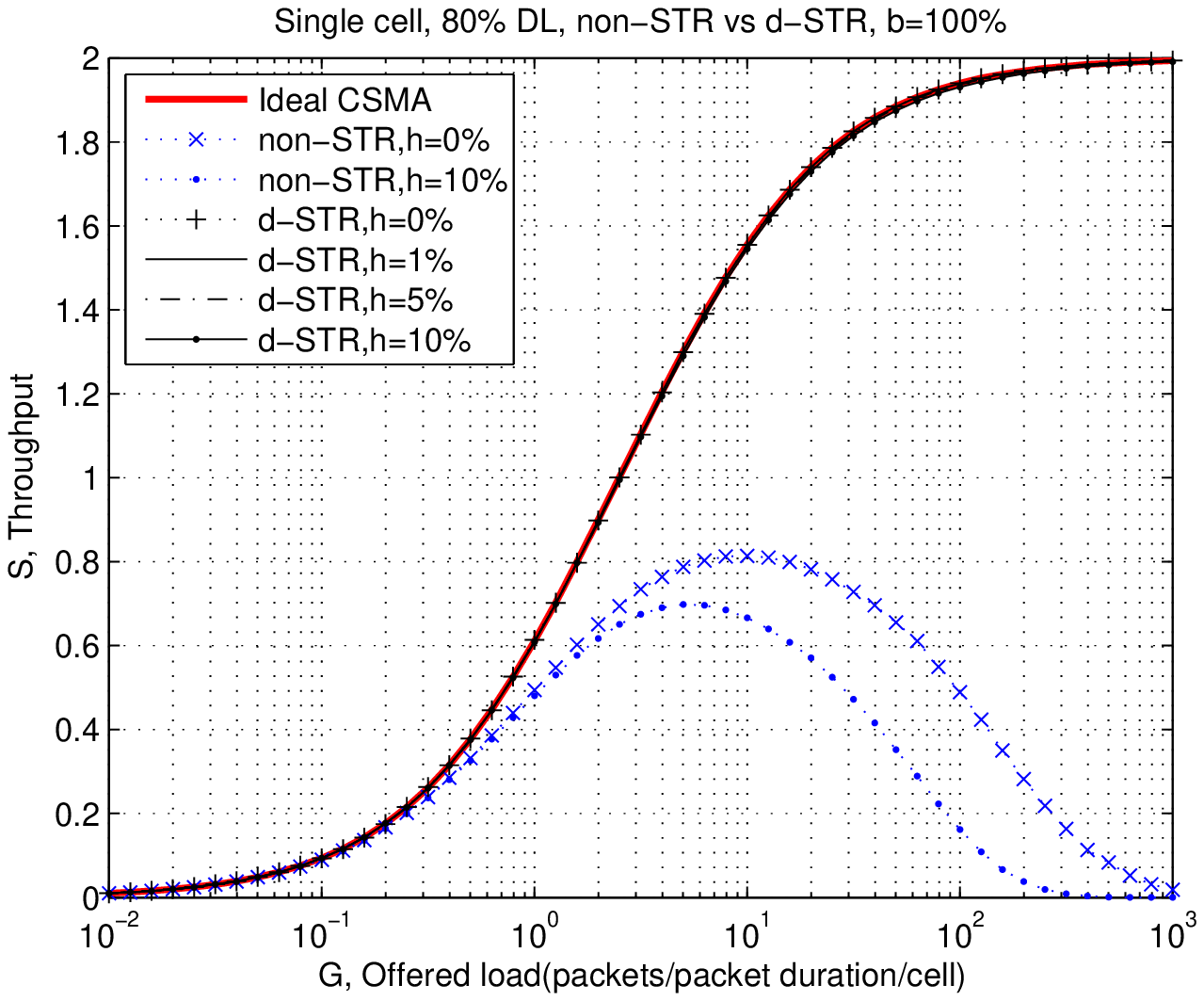}}
\hfil \subfigure[]
    {\includegraphics[width=3.0 in]{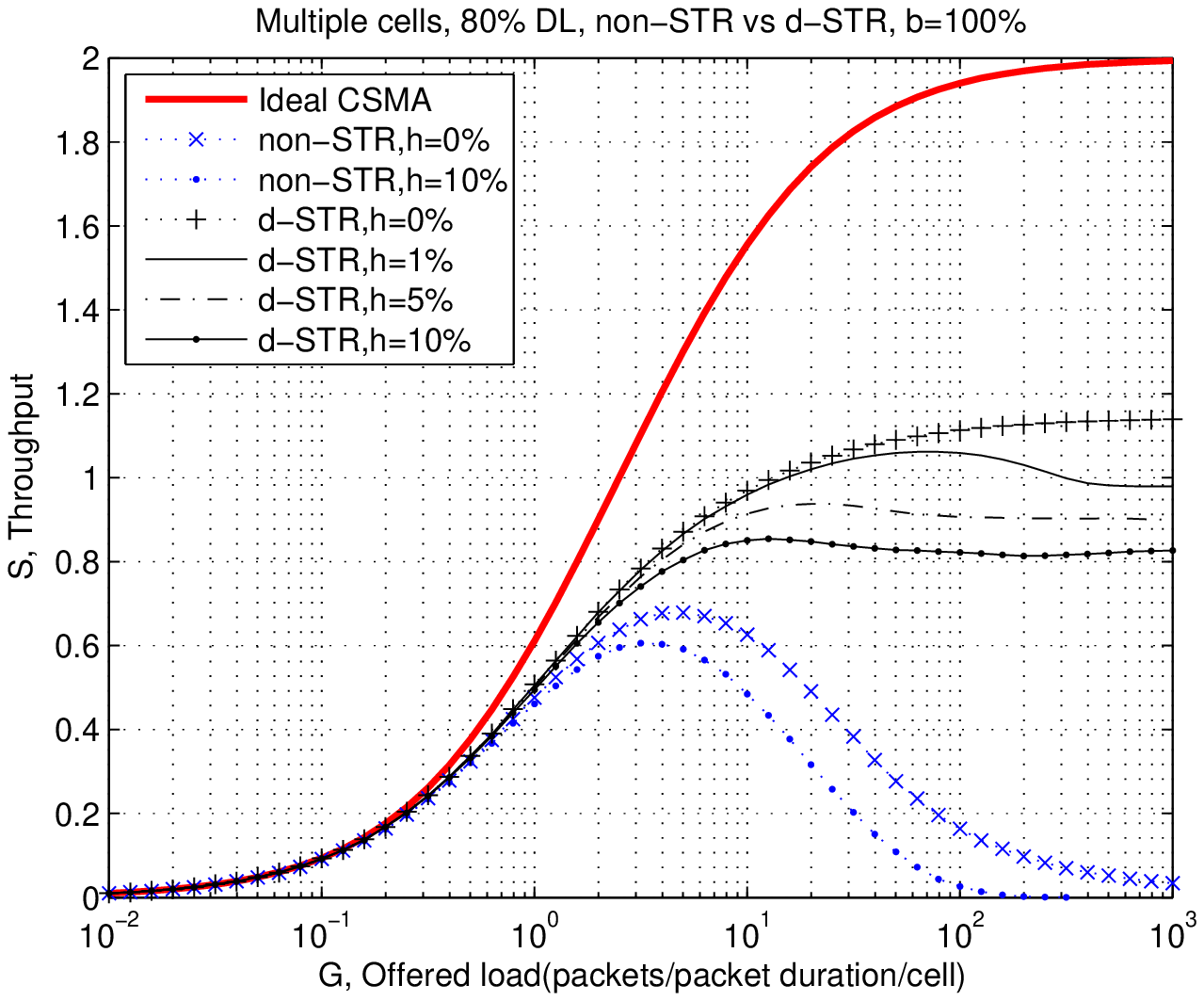}}
} \centerline{ \subfigure[]
    {\includegraphics[width=3.0 in]{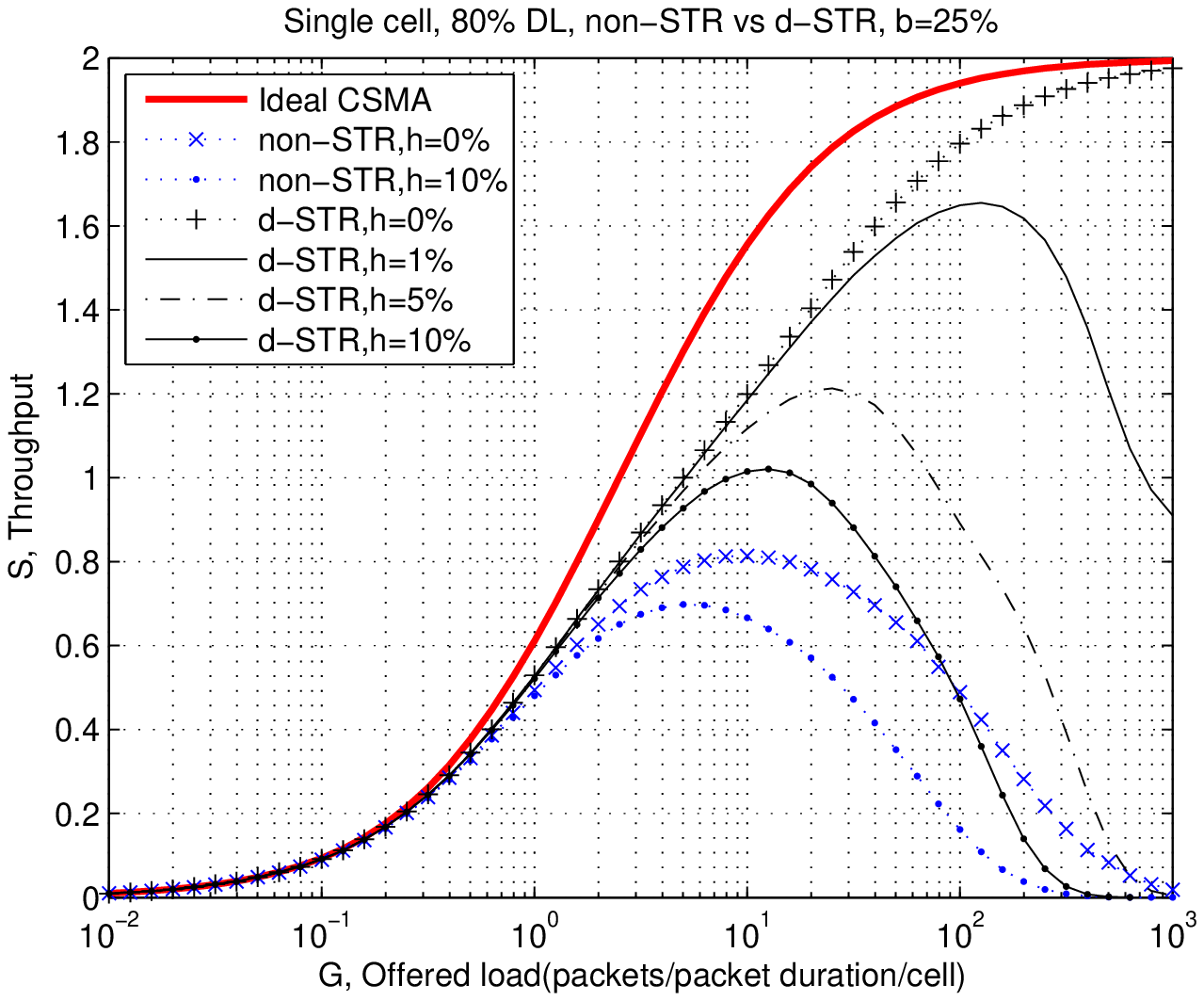}}
\hfil \subfigure[]
    {\includegraphics[width=3.0 in]{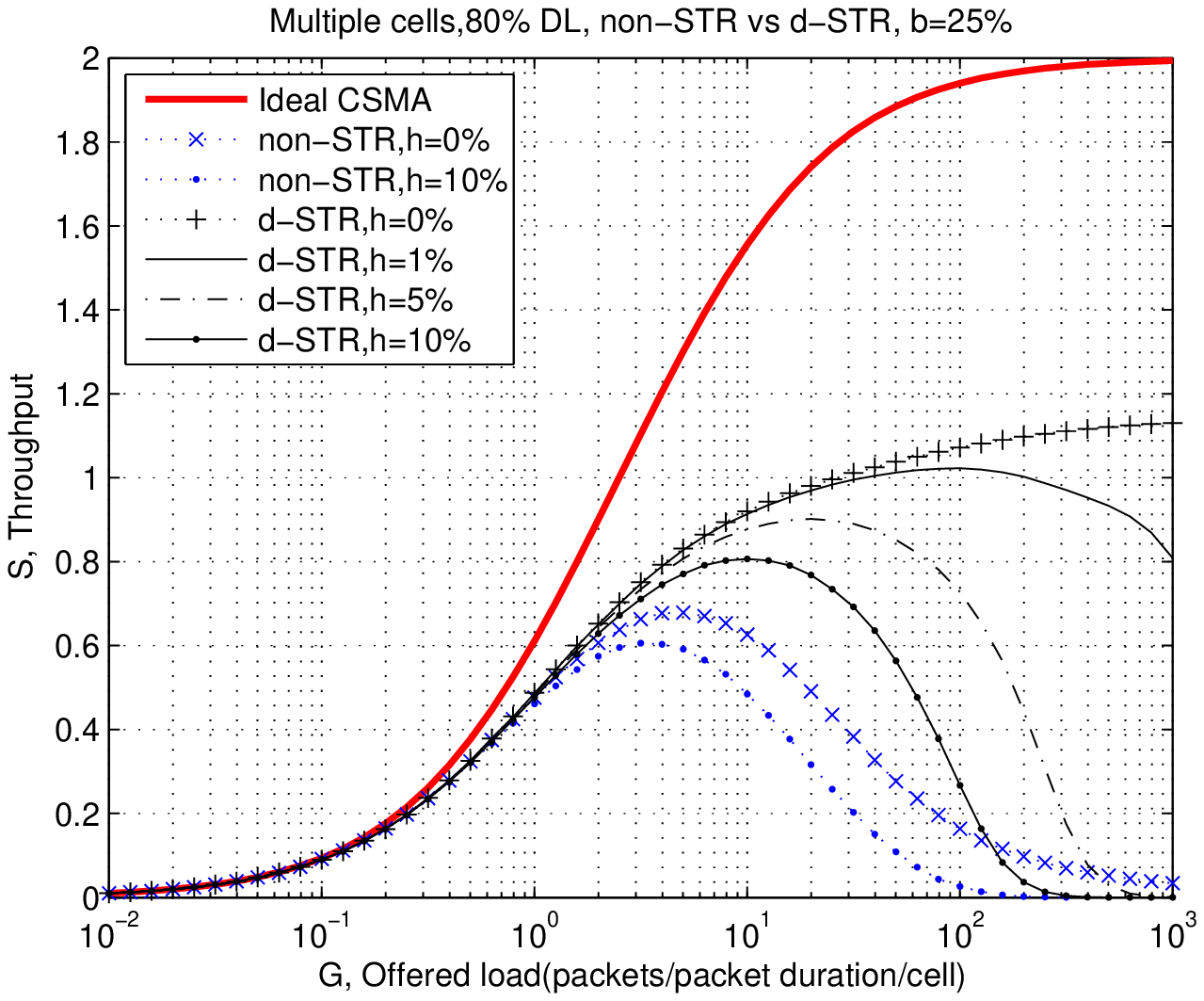}}}

\caption{Throughput comparisons in d-STR, DL traffic=80\%: (a)
single cell, $b=100\%$ (b) multiple cells, $b=100\%$ (c) single
cell, $b=25\%$ (d) multiple cells, $b=25\%$. } \label{fig_d_STR}

\end{figure*}

Simulation results for a single hop network with non-slotted
transmissions, and \textit{nonpersistent} CSMA \cite{csma} are
presented in Fig. \ref{fig_s_i_STR}. Propagation delay is assumed to
be zero in order to isolate hidden node problem from throughput
degradation due to increased contention as a result of propagation
delay. Terminals are assumed to be evenly and randomly distributed
over the entire coverage area and operate in an infrastructure mode
communicating through a central node (AP). Each packet is of
constant length and packet arrival follows Poisson process. In
practice, a collision does not necessarily result in packet error
thanks to the near-far effect and advanced receiver. However, in our
simulations we treat all collisions as packet errors. We show
throughput \textit{S} as a function of offered load \textit{G}
assuming that downlink traffic is 80\% which is favorable to non-STR
systems. In the figures the throughput of ideal CSMA (with no hidden
node) is illustrated and can be considered as an upper bound. The
throughput expression for ideal CSMA is known as
\begin{equation}
S=\frac{G}{1+G}.
\end{equation}
Note that \textit{S} is the average number of transmitted packets
without collision per transmission time and \textit{G} includes not
only newly arrived packets but also packets waiting for
retransmissions.

For multiple cell scenario we consider a 7-cell non-overlapping
lay-out with one central cell and 6 cells surrounding the central
cell. The data is collected from the center cell. The distance
between cell centers is $2r$. In this scenario when downlink and
uplink are synchronized over all cells, no collision occurs between
cells. However, perfect synchronization is impossible in random
access. In general, presence of neighboring cells increases the
collisions and causes exposed node problem. As a result, throughput
drop is unavoidable.

\begin{center}
\begin{table*}[ht]
\centering \caption{Improvement of maximum throughput over non-STR :
Downlink traffic = 80\%}
\begin{small}
\begin{tabular}{c||c|c|c|c||c|c|c|c}
  \hline
  \hline
\multirow{3}{*}{$h$}&\multicolumn{4}{c||}{Single cell}  &\multicolumn{4}{c}{Multiple cells}  \\
    \cline{2-9}
   & s-STR & i-STR & d-STR & d-STR & s-STR & i-STR & d-STR & d-STR  \\
  &   &    &(b=25\%)&(b=100\%)&  &  &(b=25\%)&(b=100\%)\\
  \hline
  \hline
  0\% & 22.8\% & 22.8\% & 143\%  & 145\% & 32.1\% & 47.2\% & 66.6\% & 67.9\% \\
  1\% & 14.0\% & 22.6\% & 107\%  & 149\% & 20.4\% & 45.3\% & 52.8\% & 58.6\%   \\
  5\% &  7.9\% & 21.2\% & 62.2\% & 166\% & 10.7\% & 37.3\% & 40.9\% & 46.6\%  \\
  10\% & 5.4\% & 18.8\% & 46.2\% & 185\% &  5.7\% & 29.6\% & 33.2\% & 41.1\% \\
\hline \hline
\end{tabular}
\end{small}
\end{table*}
\end{center}

\begin{center}
\begin{table*}[ht]
\centering \caption{Improvement of maximum throughput over non-STR :
Downlink traffic = 50\%}
\begin{small}
\begin{tabular}{c||c|c|c|c||c|c|c|c}
  \hline
  \hline
\multirow{3}{*}{$h$} &\multicolumn{4}{c||}{Single cell}  &\multicolumn{4}{c}{Multiple cells}  \\
    \cline{2-9}
   & s-STR & i-STR & d-STR & d-STR & s-STR & i-STR & d-STR & d-STR  \\
  &   &    &(b=25\%)&(b=100\%)&  &  &(b=25\%)&(b=100\%)\\
  \hline
  \hline
  0\% & 63.2\% & 63.2\% &  224\% &   226\% & 76.0\% & 89.2\% & 115\%  & 116\%\\
  1\% & 43.0\% & 60.0\% &  188\% &   231\% & 49.1\% & 83.9\% & 95.5\% & 101\%\\
  5\% & 27.2\% & 49.6\% &  119\% &   253\% & 26.9\% & 64.2\% & 73.1\% & 81.4\%\\
  10\%& 19.0\% & 39.2\% &  84.6\%&   279\% & 16.3\% & 46.7\% & 56.1\% & 67.4\%\\
\hline \hline
\end{tabular}
\end{small}
\end{table*}
\end{center}

When $h=0\%$, collisions come from only hidden nodes. However, s-STR
and i-STR systems do not have any collision thanks to the dummy
packet and, as it can be seen in Fig. \ref{fig_s_i_STR} (a) and (c),
the corresponding throughput curves match ideal CSMA. In contrast,
non-STR system suffers from hidden nodes even with $h=0\%$. This is
clear evidence that the dummy packet in s-STR and i-STR systems
completely resolves the hidden node issue. Apparently, i-STR
outperforms s-STR in both single cell and multiple cells since i-STR
releases the channel earlier when a collision is detected. In
multiple cells additional loss is observed. The loss is more
significant in non-STR systems. In fact in multiple cell case, the
dummy packet from terminal can create collision to neighboring
downlink signal. However, the gains from s-STR and i-STR in all
other cases are more dominant.

Now we further modify i-STR as follows. Instead of sending a dummy
packet, the destination node transmits a packet which contains
information. It is apparent that the throughput will be doubled
whenever the destination node sends a packet while it receives a
packet. However, due to random packet arrival, it is not always a
realistic assumption that the arrivals of transmit and receive
packet are synchronized. To be realistic we again assume Poisson
distribution for packet arrival. In this asynchronous transmission,
new packet B as shown in Fig. \ref{frame} (c) will overwrite the
dummy packet A$\sim$. The header for the implicit ACK for new packet
B can be embedded or interleaved in the packet A. This technique
will be referred to as \textit{d-STR}. In d-STR, the ideal
throughput is obtained as follows

\begin{equation}
S=\frac{pG}{1+pG}+\frac{(1-p)G}{1+(1-p)G}
\end{equation}
where $p$ represents the probability of downlink packet. Notice that
as $G$ goes to infinity, the throughput $S$ converges to 2 when
$p\neq1$ and $p\neq0$. Since asynchronous transmit and receive
packet generation is assumed, we define following probability

\begin{itemize}
\item $b$=Prob\{randomly assign d-STR mode to a link among on-going uplink (or
downlink) transmissions when newly generated packet is downlink (or
uplink) packet\}.
\end{itemize}

Fig. \ref{fig_d_STR} (a) and (b) show the throughput when $b=100\%$.
In a single cell, d-STR can achieve the ideal throughput with
$h=0\%$ and negligible loss with $h>0\%$ is observed. $b=100\%$
means new packet will be assigned to current node whenever current
on-going packet is in the opposite direction of the new packet.
However, note that during an UL transmission, when new UL packet
from hidden node is arrived during non-zero $h$, there will be a
collision. Thus, some loss is observed with $h>0\%$. Nevertheless,
this loss is negligible. Unlike in single cell, in multiple cells
the huge gain is not observed as seen in Fig. \ref{fig_d_STR} (b)
and (d). We bring attention to effectively longer packet duration in
d-STR as illustrated in Fig. \ref{frame} (c). In multiple cells, due
to this longer packet duration at neighboring cells, the chance of
simultaneous transmission and reception of packets at the same node
becomes smaller. This is because longer packet at neighboring cell
increases the chance of holding a packet due to CSMA mechanism. Due
to this increased chance of holding packets, the throughput does not
drop to zero yet as the offered load increases. By decreasing the
probability $b$ to 25\% some loss is observed as expected due to
much less chance of simultaneous transmission and reception of
packets at a given node. Note that assuming no hidden node in a
single cell, the probability $b$ can be translated to the
probability of simultaneous transmission and reception of packets
only when the load goes to infinity. Otherwise, the actual chance of
enjoying the twice gain by simultaneous transmission and reception
is much smaller than $b$. It is obvious that when load is low, the
chance of transmitting one packet while receiving another one at the
same node is small due to random arrival of packets.

Table II and III summarize the gain of STR over non-STR in maximum
throughput. It is clear that the gain is larger with smaller
downlink traffic. Not only peak throughput but also the throughput
at high load is important. As WLAN becomes more popular and the
demand grows, the load can be very high especially during busy hours
at convention centers, airports, etc. It is highly desirable to
handle high volume of traffic without too much collisions and
delays. STR systems help a lot in this heavy traffic scenarios. With
high load like $G>100$, in non-STR systems the throughput drops
quickly.

\section{System Level Simulations in Cellular Systems}

STR gain can be immediately enjoyed in any isolated link. However,
as mentioned earlier, in cellular systems, we face two unique
interferences due to multiple co-channel cells: BS-BS and UE-UE
interferences. In this section we propose methods to address these
interferences and present system level evaluation results.

\begin{table}[ht]
\caption{System level simulation parameters}
\begin{small}
\begin{tabular}{l||c|c}
  \hline
  \hline
\multirow{2}{*}{Parameters} &\multicolumn{2}{c}{Assumptions}   \\
    \cline{2-3}
   & Small Cell  & Large Cell \\
  \hline
  \hline
  Carrier frequency & \multicolumn{2}{c}{2 GHz}   \\
  \hline
  System bandwidth & \multicolumn{2}{c}{10 MHz}   \\
  \hline
  Antenna gain at BS & \multicolumn{2}{c}{14 dBi ($0.5\lambda$ width, $4\lambda$ height)}   \\
  \hline
  Number of antennas at BS & \multicolumn{2}{c}{4}   \\
   \hline
  Number of antennas at UE & \multicolumn{2}{c}{1 with 0 dBi gain}   \\
  \hline
  Transmit power at BS & 23 dBm & 46 dBm \\
  \hline
  Transmit power at UE & 0 dBm & 23 dBm \\
  \hline
  BS antenna height & 7.25 m & 25 m \\
  \hline
  UE antenna height & \multicolumn{2}{c}{1.5 m}  \\
  \hline
  Inter site distance & 122 m & 500 m \\
  \hline
  Cell radius & 71 m & 289 m \\
  \hline
  \multirow{2}{*}{Shadow fading} & \multicolumn{2}{c}{Log-normal with 8 dB}  \\
  &\multicolumn{2}{c}{standard deviation \cite{3gpp} } \\
  \hline
  \multirow{2}{*}{Fast fading} & \multicolumn{2}{c}{zero mean complex Gaussian}  \\
   & \multicolumn{2}{c}{with a variance equal to 1}  \\
  \hline
  Noise level & \multicolumn{2}{c}{-174 dBm/Hz}  \\
  \hline
  Noise figure at BS & \multicolumn{2}{c}{5 dB}  \\
  \hline
  Noise figure at UE & \multicolumn{2}{c}{9 dB}  \\
\hline \hline
\end{tabular}
\end{small}
\end{table}

\subsection{Simulation Conditions}

In our simulations a wrap-around hexagonal grid of 19 cell sites
with three sectors per site, and parameters given in Table IV are
considered. Furthermore, 3GPP simulation conditions \cite{3gpp} are
followed in general including correlation of shadow fading between
cells, etc. However, regarding fast fading, blockwise flat fading is
assumed. It is also assumed that any link between BS and UE occupies
the whole bandwidth unless mentioned otherwise.

\begin{figure}[!t]

\centering
    {\includegraphics[width=3.41 in]{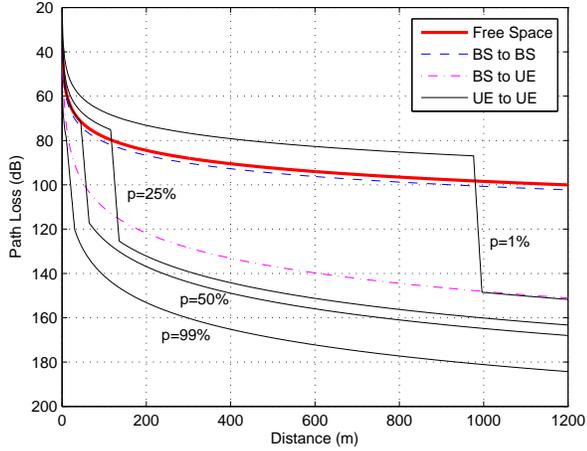}}

\caption{Path loss. \label{pl} }

\end{figure}

For UE-to-BS and BS-to-UE channels, 3GPP path loss model \cite{3gpp}
is considered with 20 dB penetration loss as shown in Fig. \ref{pl}.
BS-BS path loss model is based on coexistence study \cite{coex}.
This model assumes dual-slope LoS propagation: free-space
propagation until a certain breakpoint distance and increased
attenuation beyond the breakpoint due to diffraction/reflection
effects. The breakpoint distance depends on transmitter and receiver
heights, and wavelength $\lambda$. Notice that, as the propagation
between two base stations is LoS, no shadow fading is considered in
this model. BS-BS path loss is illustrated in Fig. \ref{pl}. Note
that the breakpoint with given parameters is beyond 1200 m and thus
it is not shown in this figure. Similar to BS-BS, UE-UE path loss
model is borrowed from coexistence study \cite{coex}. The model
consists of two separate regions of LoS and non-line-of-sight (NLoS)
with a rapid decrease in signal level between the two regions (at a
transition point). The path loss in each region depends on carrier
frequency, UE-UE distance and location percentage, $p$. Besides, $p$
determines the location of transition point. As it is observed from
Fig. \ref{pl}, with larger location percentage $p$, path loss
becomes larger. Moreover, the transition from LoS to NLoS occurs at
a smaller distance. Therefore, with smaller location percentage $p$,
UEs suffer more from UE-UE interference while larger location
percentage $p$ results in smaller UE-UE interference in the system.
In addition to the described UE-UE path loss, a penetration loss of
20 dB is added with 50\% probability to each UE-UE link to model
in-building propagation.

\subsection{BS-BS Interference}

\begin{figure}[!t]

\centering {\subfigure[] {
    {\includegraphics[width=3.41 in]{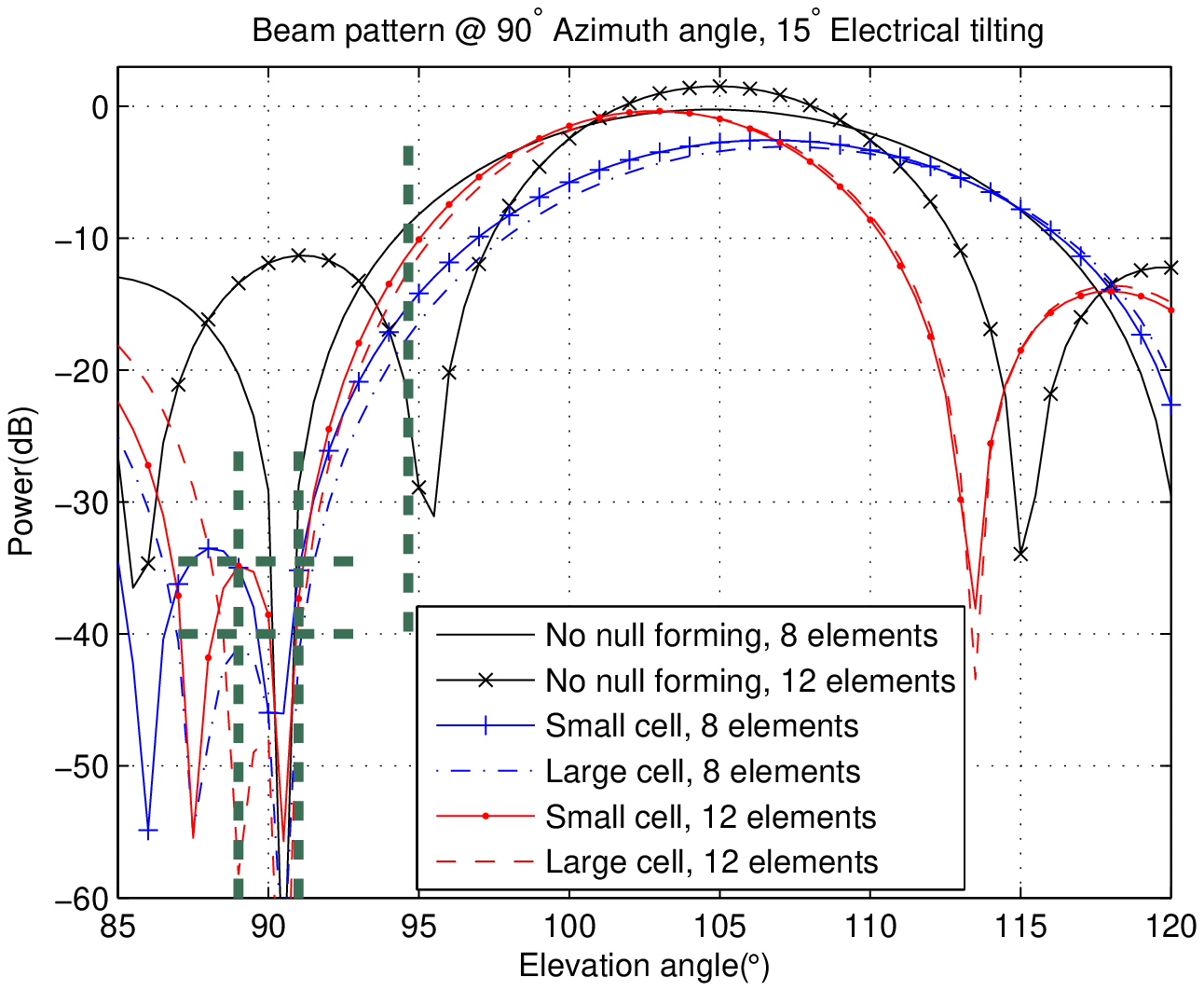}} }
\subfigure[] {
    {\includegraphics[width=3.41 in]{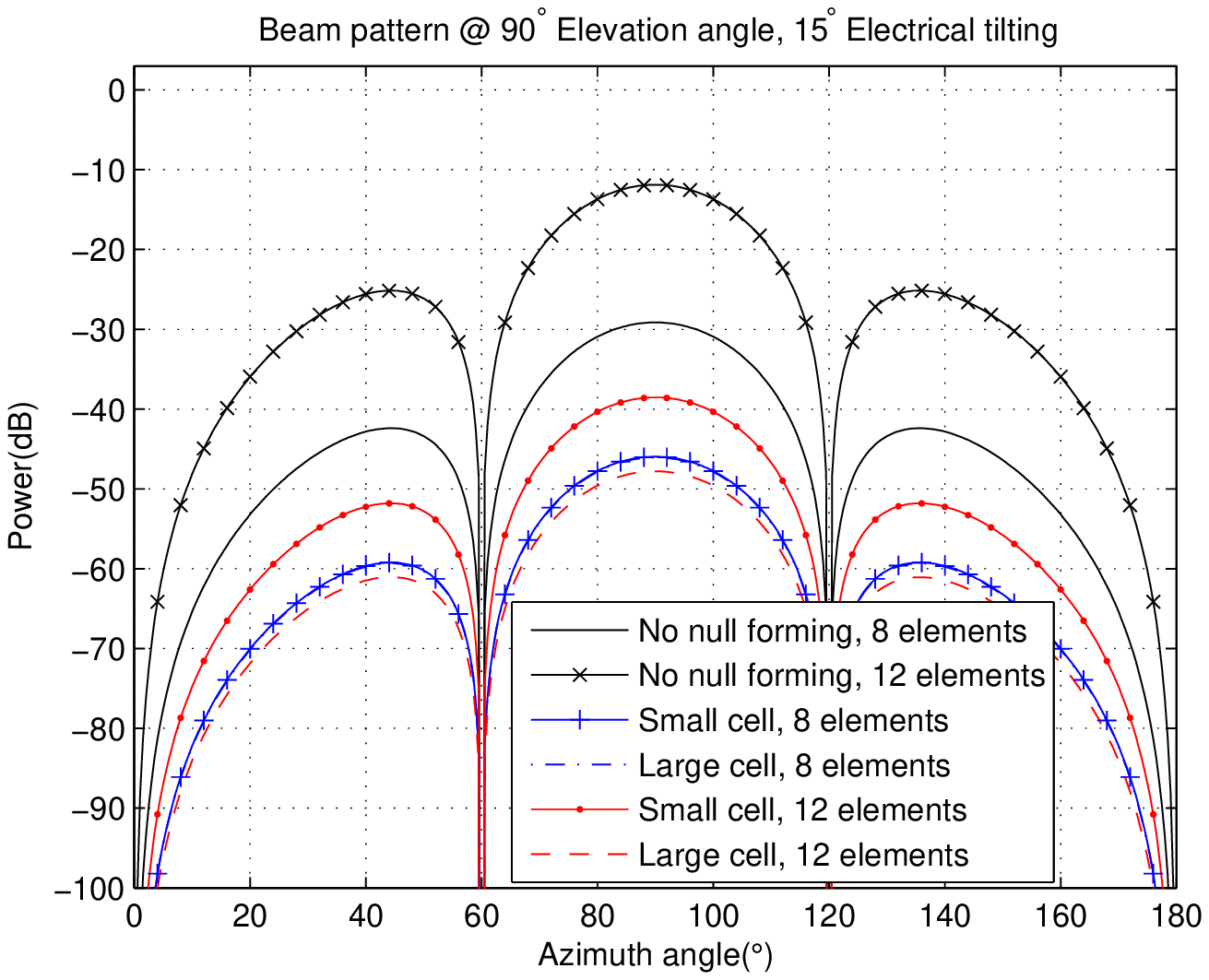}} } }

\caption{Beam patterns with and without null forming: (a) Elevation
angle (b) Azimuth angle. \label{beam} }

\end{figure}

As discussed in Section I, with STR at BS, BS-BS interference is an
extremely serious issue in cellular systems and overwhelms weak UL
signal. Hence, serious loss of UL capacity is observed unless this
interference is greatly reduced. In order to cope with BS-BS
interference, we suggest a null forming in elevation angle at BS
antennas. As tilting is employed at each BS and every BS has similar
height, by simply creating nulls at the vicinity of $90^{\circ}$ in
elevation angle, BS-BS interference can be avoided. First, we split
each $4 \lambda$ height antenna to eight $1/2 \lambda$ elements
vertically stacked. Notice that as the effective sizes of the
antennas in the two configurations are equivalent, antenna gains
will be unchanged. Since we have 4 elements horizontally, we have
two dimensional $4 \times 8$ antenna array. Assuming that $M \times
N$ planar array is located at \textit{x-z} plane facing toward
positive \textit{y}-axis with $1/2 \lambda$ spacing of all elements,
we have the following beam pattern

\begin{align} \label{antpattern}
&A(\theta, \phi)= \nonumber\\
&\left |\frac{1}{MN} \sum_{m=0}^{M-1} \sum_{n=0}^{N-1} w_{m,n}
e^{jm\pi \sin (\theta) \cos(\phi)} e^{jn\pi \cos (\theta) }  \right
|^2 A_{0}(\theta, \phi)
\end{align}
where $w_{m,n}$ is the weight for $(m,n)$-th element, $M$ is the
number of horizontal elements, $N$ is the number of vertical
elements, $\theta$ is elevation angle measured from positive
\textit{z}-axis, $\phi$ is azimuth angle measured from positive
\textit{x}-axis toward positive \textit{y}-axis in \textit{x-y}
plane, and $A_{0}(\theta, \phi)$ is the beam pattern of $1/2
\lambda$ height and $1/2 \lambda$ width rectangular antenna element.
By ignoring antenna coupling, etc., the following ideal beam pattern
is obtained \footnote{Although not shown in this paper, system level
simulations results showed no significant difference between this
pattern and the one considered in 3GPP evaluations.}

\begin{equation}
A_{0}(\theta, \phi)= \left | Sinc\left ( \frac{ \cos (\theta)} {2}
\right ) Sinc\left ( \frac {\sin (\theta) \cos(\phi)} {2} \right)
\right |^2.
\end{equation}
The above pattern \eqref{antpattern} is used for $0 \le \phi, \theta
\le \pi$ and scaled down by 25 dB for the backlobe ($\phi>\pi$).
Although we can make a null at an arbitrary azimuth and elevation
angle, we create a null at particular elevation angle over the
entire range of azimuth angle for simplicity. Therefore, the weight
$w_{m,n}$ can be split into horizontal and vertical weights as shown
below

\begin{equation}
w_{m,n}=w_m^h w_n^v
\end{equation}
where superscript $(\cdot)^h$ and $(\cdot)^v$ represent horizontal
and vertical weight, respectively. The horizontal weights $\{ w_m^h
\}$ come from conventional closed-loop or open-loop MIMO techniques.
In this paper without loss of generality we assume $w_m^h=1$. With
$15^{\circ}$ electrical down tilting, the vertical weights become

\begin{equation}
w_n^v = e^{-j n\pi \cos (105 \pi /180 )}.
\end{equation}

In order to create nulls MMSE beamforming is used. Then, the
following row vector creates nulls toward $\{\theta_k, k\neq0\}$
while maintaining main beam direction $\theta_0$

\begin{equation}
\textbf{\textit{w}}^v = \textbf{\textit{a}}^H(\theta_0) \left
(\textbf{\textit{a}}(\theta_0) \textbf{\textit{a}}^H(\theta_0) +
\sum_{k\neq 0} \textbf{\textit{a}}(\theta_k)
\textbf{\textit{a}}^H(\theta_k) +\epsilon \textbf{\textit{I}} \right
)^{-1}
\end{equation}
where $\textbf{\textit{a}}(\theta_0)= [1~ e^{j \pi \cos (\theta_0)}~
\cdots e^{j (N-1)\pi \cos (\theta_0)}]^T $ is an array vector toward
main beam direction $\theta_0$, $\epsilon$ controls the depth of
nulls, and superscript $(\cdot)^H$ represents complex conjugate
transpose. We normalize the weight vector by maximum magnitude of
its elements so that the magnitude of any element does not exceed 1

\begin{equation}
\textbf{\textit{w}}^v = \textbf{\textit{w}}^v/\max_n |w_n^v|.
\end{equation}
The above normalization is more practical in transmit beamforming
than finite norm of the weight vector although some loss of power is
expected in transmission mode. The null forming is applied to both
transmit and receive beam, which will relax the requirement on the
depth of nulls. We create wide nulls from $89^\circ$ to $91^\circ$
for 25 m high BS antennas to allow variation of antenna heights from
16.3 m to 33.7 m at neighboring BSs located at 500 m distance.

Fig. \ref{beam} (a) illustrates beam patterns with and without null
forming as a function of elevation angle at $90^\circ$ azimuth
angle. With $15^\circ$ electrical tilting, the beam pattern of 12
antenna elements without null forming has high gain at around
$90^\circ$ elevation angle. However, the beam of 8 antenna elements
has a null at $90.5^\circ$. Hence, by changing the tilting angle to
$14.5^\circ$, the null points to $90^\circ$. Unfortunately, the null
is not wide enough to accommodate the variation of the antenna
heights. Hence, for wider nulls, we apply the proposed technique to
both 8 and 12 antenna elements. Two vertical lines around $90^\circ$
in the Fig. \ref{beam} (a) show the range of nulls and two
horizontal lines represent the targets on the depth of nulls for the
small and large cell. Since the transmit power in the small cell is
23 dB smaller while the path loss from the first tier BSs is 12.3 dB
smaller, the requirement on the depth of nulls is looser in the
small cell. Vertical line at $94.65^\circ$ represents the elevation
angle pointing to cell edge UEs. Due to null forming we observe some
loss in signal power especially at cell edge. One option to fix this
is to reduce the cell size. Since the path loss exponent of BS-BS
channel is smaller than that of BS-UE channel, reducing the cell
size will help to improve cell edge performance. Also by increasing
the number of antenna elements we can overcome the loss to certain
extent as seen in the Fig. \ref{beam} (a). From $90^\circ$ to
$94.65^\circ$ it is noticeable that the gain of beam due to null
forming is smaller than the gain of the beam pattern without null
forming. Thus, it reduces the co-channel interference to and from
neighboring cells. Due to the reduction in interference power,
although the signal power is reduced due to null forming, in
interference limited systems not much loss of capacity is observed
at cell edge, and a gain in cell center is possible. Fig. \ref{beam}
(b) shows the beam patterns as a function of azimuth angle at
$90^\circ$ elevation angle. It is clear that over entire range of
horizontal angle the beams with null forming meet the requirements
on the depth of nulls. Due to the assumption of 25 dB backlobe
rejection, null forming and 40 dB path loss at 1 m, BS-BS
interferences between sectors at the same site is negligible as
well.

\begin{figure}[!t]

\centering
    {\includegraphics[width=3.41 in]{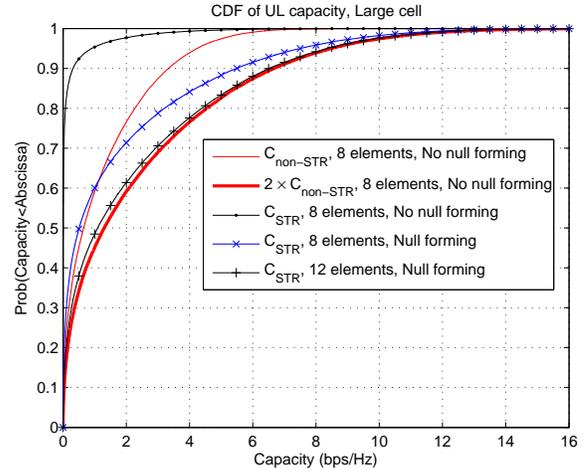}}

\caption{CDF of UL capacity in large cell. \label{BSBSLarge} }

\end{figure}

Fig. \ref{BSBSLarge} depicts the CDFs of UL capacity with various
weight vectors in the large cell. The large cell is worse than the
small cell regarding BS-BS interference especially in cell edge
performance. Thus, we show the CDFs of the large cell. In order to
simulate various antenna heights, we randomly choose BS-BS angle
between $89^\circ$ and $91^\circ$. Throughout this paper we treat
non-STR system as a reuse 2 system since DL and UL require separate
non-overlapping resources. Therefore, we divide non-STR capacity by
two. As seen in Fig. \ref{BSBSLarge}, due to BS-BS interference, STR
exhibits serious loss of capacity unless null forming is employed.
When null forming is employed, the CDF of STR capacity approaches
twice the non-STR capacity. Although due to null forming some loss
in signal power is observed, this loss can be overcome by reducing
the cell size without changing the transmit power and antenna gain,
or by increasing the number of antenna elements.

\subsection{UE-UE Interference}

\begin{figure}[!t]

\centering
    {\includegraphics[width=3.41 in]{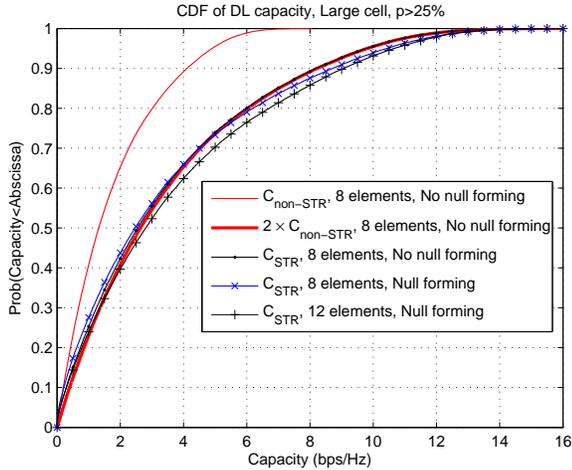}}

\caption{CDF of DL capacity in large cell. \label{LargeDL} }
\end{figure}

The significance of UE-UE interference depends on its power relative
to BS-UE interference power. With random choice of $p$ in each UE-UE
link in the range of $25\%<p<99\%$, we observed that UE-UE
interference is more severe than BS-UE interference in the small
cell but weaker in the large cell. Hence, we can ignore UE-UE
interference in the large cell since BS-UE interference is more
dominant. Fig. \ref{LargeDL} demonstrates the gain of DL capacity
including UE-UE interference impact when $25\%<p<99\%$. Clearly, STR
achieves twice of non-STR capacity almost everywhere. Note that the
STR capacity with null forming can be better than twice of non-STR
capacity. This comes from the fact that null forming reduces
co-channel interference. Thus, in the interference limited system,
when null forming is employed, the capacity of STR can be more than
double of non-STR capacity except at cell edge. The cell edge
performance can be improved by adjusting the cell size to compensate
the signal power loss due to null forming. When UE-UE channel model
is favorable, STR combined with null forming results in significant
gains in both DL and UL.

However, it is obvious that when UE-UE path loss is smaller than
BS-UE path loss, UE-UE interference will be more dominant than BS-UE
interference. In the small cell the distance between UEs becomes
smaller as the cell radius is 70.6 m. As it can be observed from
Fig. \ref{pl}, when $p=25\%$, in distances up to 115 m, UE-UE
channel is LoS and its path loss is always much less than the one
for BS-UE. For example at 115 m, UE-UE path loss is 75 dB while
BS-UE path loss is 113 dB even though BS has higher antenna height
than  UE. Thus, UE-UE interference is more dominant in small cell in
this setting. In practice there is a correlation between BS-UE and
UE-UE channel model. When UE-UE channel has high chance of LoS,
BS-UE may also have high chance of LoS. However, due to lack of
channel models that include such correlation, we study the system
level performances with various values of $p$ to simulate different
environments.

It is apparent that coordination over multiple cells can reduce
UE-UE interference by joint scheduling. For example, when two UEs
see strong UE-UE interference from each other, scheduler can
schedule them in orthogonal resources. There is no doubt that
through the coordination, UE-UE interference can be avoided. The
UE-UE interference from neighboring sectors at sector boundary is
also considerably significant in reuse 1. Since a coordination
between sectors from the same site is relatively easier, the
coordination in the same site is quite feasible. In addition
aggregating multiple BSs at one physical signal processing center is
being developed called Cloud Radio Access Network (C-RAN) which
makes the coordination easier. Recently, a coordination among
multiple BSs has been defined in 3GPP \cite{comp}. By exchanging
location information of UEs among neighboring BSs, to some degree we
can avoid UE-UE interference as well. In general, some form of
coordination will be made available in near future.

\begin{figure}[!t]

\centering
    {\includegraphics[scale=1.3]{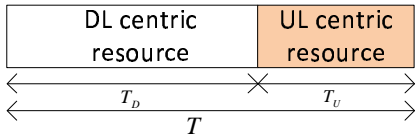}}
\caption{Basic resource block structure for UE-UE interference
reduction. \label{frame_UE_UE} }

\end{figure}

However, in this work, for the sake of simplicity we focus on a
technique which does not require any coordination. Under this
constraint, there is no other way than turning off some UEs' UL
transmissions which create severe UE-UE interferences to other UEs.
Thus, DL capacity can be improved at the cost of UL capacity. For
this end, we start off with half duplex. A basic resource block is
suggested in Fig. \ref{frame_UE_UE} where the resource block is
split into two parts. One for DL centric resource with size of $T_D$
and the other for UL centric resource with size of $T_U$. Note that
a packet is transmitted using multiple of the basic resource blocks.
During the DL centric resource, UEs which can benefit from
additional UL transmission by STR are selected for STR operation.
This additional UL will be a gain in UL capacity but it creates
UE-UE interferences to the DL signals in other cells. Obviously the
additional UL gain should be larger than the DL loss. Due to the
assumption of no coordination, we simply check DL signal
power-to-interference and noise power ratio (SINR) in order to
choose STR enabled UEs during the DL resource. Likewise, over the UL
duration, we search UEs which can provide net gain by enabling DL
transmission using STR. The DL signal for the STR UE will not
interfere with UL signals due to null forming at BS. However, it
creates conventional DL co-channel interference to other DL signals
in neighboring cells.

Unlike DL co-channel interference from BS to UE, UE-UE interference
is dependent on the locations of other UEs. Thus, information on
which UEs are scheduled in other BSs is necessary. We need to
address how to obtain DL SINR that includes UE-UE interferences
without knowing the scheduling decisions of neighboring BSs. We
suggest the following: first, each BS schedules UE. Then, the
scheduled UE sends UL pilots without sending data yet. Using the UL
Pilots, the UE measures the interferences from other UEs in
co-channel BSs. Then, the UE calculates DL SINR assuming STR
operation in all UEs. Afterwards, each UE reports the DL SINR to the
home BS. If the DL SINR is lower than a certain threshold, UE is
scheduled to operate in non-STR mode in successive resource.
Otherwise, it is scheduled to operate in STR mode. The thresholds
for DL and UL centric resources may be different in general. In
addition, by optimizing the amount of resources assigned to DL and
UL centric zones, capacity can be further improved. Thus, we have
three dimensional optimization problem. The proposed resource block
structure provides good flexibility between DL and UL capacity.

\begin{center}
\begin{table*}[ht]
\centering \caption{Improvement of capacity over non-STR with 8
antennas and no null forming: Large cell}
\begin{small}
\begin{tabular}{c||c|c|c|c|c|c||c|c}
  \hline
  \hline
Number &\multicolumn{6}{c||}{DL}  &\multicolumn{2}{c}{UL}  \\
    \cline{2-9}
 of  & \multicolumn{2}{c|}{$p>$0.25} & \multicolumn{2}{c|}{$p>$0.37} & \multicolumn{2}{c||}{$p>$0.5} & \multicolumn{2}{c}{}  \\
\cline{2-9}
 Antennas & Mean & Edge & Mean & Edge & Mean & Edge & Mean & Edge\\
  \hline
  \hline
  8 & 100.6\% &-35.0\% & 103.0\% & 0.2\% & 103.8\% & 17.5\% & 42.8\% &-50.0\% \\
 12 & 116.8\% &  2.2\% & 119.3\% & 46.7\%& 120.1\% & 66.7\% & 90.9\% & 32.8\%   \\
\hline \hline
\end{tabular}
\end{small}
\end{table*}
\end{center}

\begin{center}
\begin{table*}[ht]
\centering \caption{Improvement of capacity over non-STR with 8
antennas and no null forming: Small cell}
\begin{small}
\begin{tabular}{c||c|c|c|c||c|c}
  \hline
  \hline
Number &\multicolumn{4}{c||}{DL}  &\multicolumn{2}{c}{UL}  \\
    \cline{2-7}
 of   & \multicolumn{2}{c|}{$p>$0.25} & \multicolumn{2}{c||}{$p>$0.37} & \multicolumn{2}{c}{}\\
\cline{2-7}
 Antennas & Mean & Edge & Mean & Edge & Mean & Edge \\
  \hline
  \hline
  8 & 37.7\% & -92.6\% & 69.8\% & -80.9\% & 55.8\% & -27.5\%  \\
 12 & 55.3\% & -82.4\% & 84.8\% & -59.7\% & 83.5\% &  36.2\%  \\
\hline \hline
\end{tabular}
\end{small}
\end{table*}
\end{center}

\begin{center}
\begin{table*}[ht]
\centering \caption{Improvement of capacity over non-STR with 8
antennas and no null forming: Small cell, using the resource block
structure}
\begin{small}
\begin{tabular}{c||c|c|c|c||c|c|c|c}
  \hline
  \hline
Number &\multicolumn{4}{c||}{DL}  &\multicolumn{4}{c}{UL}  \\
    \cline{2-9}
 of   & \multicolumn{2}{c|}{$p>$0.25} & \multicolumn{2}{c||}{$p>$0.37} & \multicolumn{2}{c|}{$p>$0.25} & \multicolumn{2}{c}{$p>$0.37}\\
\cline{2-9}
 Antennas & Mean & Edge & Mean & Edge & Mean & Edge & Mean & Edge \\
  \hline
  \hline
  8 & 64.6\% & 2.0\% & 81.1\% & -0.9\% & 0.2\% & -53.2\% & 27.9\% & -45.7\% \\
 12 & 71.3\% & 6.9\% & 92.1\% & 17.1\% &44.6\% &   8.4\% & 63.7\% & 16.3\% \\
\hline \hline
\end{tabular}
\end{small}
\end{table*}
\end{center}

However, one difficulty in solving this optimization problem is the
lack of a well-known cost function that can capture gains in cell
edge and cell average in both DL and UL. In general different cost
functions can be considered depending on the application and the
required gain in each parameter. Here, our objective is to simply
maximize the sum of all gains \emph{i.e.}, DL cell edge gain + UL
cell edge gain + DL cell average gain + UL cell average gain.

Table V-VII show gains in average and 5\% cell edge capacity of STR
with null forming over non-STR with 8 antennas without null forming.
First, consider the large cell. As discussed earlier in this case,
UE-UE interference is not a serious problem. It can be seen from
Table V that more than 100\% of gain in DL cell average capacity can
be achieved by STR with null forming in both 8 and 12 antennas and
decent gain in cell edge with 12 antennas. The gains beyond 100\% in
cell average come from the reduction of DL co-channel interference
due to null forming. However, since UL is less interference limited,
the gain is smaller. The cell edge capacity is not close to twice of
non-STR capacity since the signal power loss is unavoidable at cell
edge due to null forming.

Next, let us consider the case where UE-UE interference is more
dominant than DL co-channel interference \emph{i.e.} small cell with
small $p$. As it can be observed from Table VI, without addressing
UE-UE interference, cell edge users experience serious loss in DL
capacity. Nevertheless, by employing the proposed resource block
structure and optimizing above cost function, positive gains can be
achieved with STR and 12 antennas by sacrificing UL capacity as
shown in Table VII.

Although we assumed that each UE uses the whole bandwidth for time
division multiplexing (TDM), it is possible to allocate fraction of
bandwidth to each UE for frequency division multiplexing (FDM).
Then, UE-UE interference will be larger unless the power spectral
density is maintained the same. In addition, due to proximate UEs,
in the same sector higher resolution ADC will be needed. Due to path
loss, the saturation at LNA can be avoided.

\section{Conclusions}

We have suggested an echo cancellation technique which can be
implemented in analogue domain and demonstrated sufficient
suppression of echo before LNA. The technique is robust to RF
impairments exhibiting outstanding performance without requiring
additional antennas.

STR can be employed in CSMA networks. We showed that STR can reduce
the hidden node problem and the suggested protocols improve
throughput in both single and multiple cells. The application of STR
to cellular systems creates BS-BS and UE-UE interferences. We have
provided a solution for the complete cancellation of BS-BS
interference. For UE-UE interference, we studied non-cooperative
method.

\section*{Appendix A}

\setcounter{equation}{0}
\renewcommand{\theequation}{A.\arabic{equation}}

First, we define normalized correlation as

\begin{align}
\textbf{\textit{r}}_{{}_{Y\textbf{\textit{X}}}}&=\frac{1}{P}E\{
X(t-\tau) [X^*(t-\tau_1)~
X^*(t-\tau_2)]\} \nonumber\\
&=[Sinc(B(\tau-\tau_1))~Sinc(B(\tau-\tau_2))]
\end{align}
where $P$ is the average power of $X(t)$. Similarly, define
normalized auto-correlation as

\begin{align}
&\textbf{\textit{r}}_{{}_{\textbf{\textit{XX}}}}=\frac{1}{P} \nonumber \\
&\left[ {{\begin{array}{*{2}c}
E\{ X(t-\tau_1) X^*(t-\tau_1) \}  & E\{ X(t-\tau_1) X^*(t-\tau_2) \} \\
E\{ X(t-\tau_2) X^*(t-\tau_1) \}  & E\{ X(t-\tau_2) X^*(t-\tau_1) \} \\
\end{array} }}
\right] \nonumber\\
=&\left[ {{\begin{array}{*{2}c}
1                       & Sinc(B(\tau_1-\tau_2)) \\
Sinc(B(\tau_2-\tau_1))  & 1\\
\end{array} }}
\right].
\end{align}

\noindent Then, the cross correlation vector can be written as

\begin{equation}
\textbf{\textit{R}}_{YX}=E\{ Y(t) \textbf{\textit{X}}^H(t)\} = g
e^{j\omega \tau} P \textbf{\textit{r}}_{{}_{Y\textbf{\textit{X}}}}
\textbf{\textit{D}}
\end{equation}
where $\textbf{\textit{D}}$ is a diagonal matrix defined as

\begin{equation}\label{D}
\textbf{\textit{D}}=\left[ {{\begin{array}{*{2}c}
 e^{-j\omega \tau_1}  & 0 \\
0                     & e^{-j\omega \tau_2}  \\
\end{array} }}
\right].
\end{equation}

\noindent Likewise, the auto-correlation matrix can be expressed as

\begin{equation}
\textbf{\textit{R}}_{XX}=E\{ \textbf{\textit{X}}(t)
\textbf{\textit{X}}^H(t)\} = P
\textbf{\textit{D}}^*\textbf{\textit{r}}_{{}_{\textbf{\textit{XX}}}}
\textbf{\textit{D}}. \label{auto}
\end{equation}

The weight vector for the Wiener solution is obtained by

\begin{equation}
\textbf{\textit{W}}^T=\textbf{\textit{R}}_{YX}\textbf{\textit{R}}_{XX}^{-1}
=ge^{j\omega
\tau}\textbf{\textit{r}}_{{}_{Y\textbf{\textit{X}}}}\textbf{\textit{r}}_{{}_{\textbf{\textit{XX}}}}^{-1}\textbf{\textit{D}}.
\end{equation}

\noindent After manipulations, the residual echo power defined in
the denominator of \eqref{res} is in the form of

\begin{equation}\label{resEcho}
E \{ |g X(t-\tau) e^{j\omega \tau} - \textbf{\textit{W}}^T
\textbf{\textit{X}} (t) |^2 \}=g^2 P-g^2 P
\textbf{\textit{r}}_{{}_{Y\textbf{\textit{X}}}}\textbf{\textit{r}}_{{}_{\textbf{\textit{XX}}}}^{-1}\textbf{\textit{r}}_{{}_{Y\textbf{\textit{X}}}}^H.
\end{equation}

Finally, the suppression level of echo power \eqref{sup} is
calculated from

\begin{equation}
\text{Suppression}=\frac{1}{1-
\textbf{\textit{r}}_{{}_{Y\textbf{\textit{X}}}}\textbf{\textit{r}}_{{}_{\textbf{\textit{XX}}}}^{-1}\textbf{\textit{r}}_{{}_{Y\textbf{\textit{X}}}}^H}.
\end{equation}

\section*{Appendix B}

\setcounter{equation}{0}
\renewcommand{\theequation}{B.\arabic{equation}}

In order to give some insight on how phase noise affects the
performance, without loss of generality, the following simplified
model is considered. We consider one echo and assume ideal knowledge
of the delay. With one tap echo canceller and zero valued weight
initially, we have

\begin{equation}
Z(t)=HX(t)
\end{equation}
where $H=|H|e^{j \theta_h}$ is the echo channel response. The echo
channel estimation for in-phase which drives the weight is

\begin{equation}\label{errsig}
Re\{X^*(t)Z(t)\}=|H||X(t)|^2 cos(\theta_h).
\end{equation}
Ideally, the weight should converge to

\begin{equation}
w=|H| cos(\theta_h).
\end{equation}
However, the echo channel estimation (\ref{errsig}) is far from $|H|
cos(\theta_h)$ unless it is divided by $|X(t)|^2$. In addition, the
echo channel estimation will be corrupted by noise and various
impairments. Note also that random OFDM signal $X(t)$ has large
peak-to-average power ratio (PAPR). Hence, in adaptive echo
canceller, the echo channel estimation is low-pass filtered by an
integrator with very small step-size $\mu$ as the echo channel
estimation is quite noisy. Thus, the estimation error in echo
channel estimation does not directly appear in the weight. As far as
the sign of $Re\{X^*(t)Z(t)\}$ is equal to the sign of true echo
channel $|H| cos(\theta_h)$, the weight will be gradually updated
toward the true echo channel until the power of $Z(t)$ is small
enough. Due to this nature, even if the echo channel estimation is
not perfect and corrupted by impairments and noise, the adaptive
echo canceller can converge.

\begin{figure*}[!t]
\setlength{\arraycolsep}{0.0em}
\begin{eqnarray} \label{ErrWPN}
Re\{ \tilde{X}^*(t) \tilde{Z}(t) e^{j\Delta} \}=
\frac{|H||X(t)|^2}{2} && \left \{ \left [ \cos(\theta_h) \left\{
\begin{array}{l}
         g_{x,i}g_{z,i}(\cos(\delta_{i-i}+\varepsilon_{x-z}(t))+\cos(2\theta_x(t)-\delta_{i+i}-\varepsilon_{x+z}(t)))+ \\
         g_{x,q}g_{z,q}(\cos(\delta_{q-q}+\varepsilon_{x-z}(t))-\cos(2\theta_x(t)-\delta_{q+q}-\varepsilon_{x+z}(t)))\end{array} \right
         \}  \right. \right. \nonumber \\
&& {-}\: \left. \sin(\theta_h) \left\{ \begin{array}{l}
         g_{x,i}g_{z,i}(\sin(\delta_{i-i}+\varepsilon_{x-z}(t))+\sin(2\theta_x(t)-\delta_{i+i}-\varepsilon_{x+z}(t)))+ \\
         g_{x,q}g_{z,q}(\sin(\delta_{q-q}+\varepsilon_{x-z}(t))-\sin(2\theta_x(t)-\delta_{q+q}-\varepsilon_{x+z}(t)))\end{array} \right
         \} \right ] \cos(\Delta)  \nonumber \\
&& {-}\! \left [ \cos(\theta_h) \left\{ \begin{array}{l}
         g_{x,i}g_{z,q}(\sin(\delta_{i-q}+\varepsilon_{x-z}(t))+\sin(2\theta_x(t)-\delta_{i+q}-\varepsilon_{x+z}(t)))+ \\
         g_{x,q}g_{z,i}(\sin(\delta_{q-i}+\varepsilon_{x-z}(t))-\sin(2\theta_x(t)-\delta_{q+i}-\varepsilon_{x+z}(t)))\end{array} \right
         \}  \right. \nonumber \\
&& {+}\: \left. \left. \sin(\theta_h) \left\{ \begin{array}{l}
         g_{x,i}g_{z,q}(\cos(\delta_{i-q}+\varepsilon_{x-z}(t))+\cos(2\theta_x(t)-\delta_{i+q}-\varepsilon_{x+z}(t)))+ \\
         g_{x,q}g_{z,i}(\cos(\delta_{q-i}+\varepsilon_{x-z}(t))-\cos(2\theta_x(t)-\delta_{q+i}-\varepsilon_{x+z}(t)))\end{array} \right
         \} \right ] \sin(\Delta) \right \}. \nonumber \\
\end{eqnarray}
\begin{eqnarray} \label{EErrWPN}
&& E\{ Re\{ \tilde{X}^*(t) \tilde{Z}(t) e^{j\Delta}  \} \}=
\frac{e^{-\sigma^2}|H|P}{2}
\nonumber \\
&& \;\; \left \{ \left [ \cos(\theta_h) \left\{
         g_{x,i}g_{z,i}\cos(\delta_{i-i})+ g_{x,q}g_{z,q}\cos(\delta_{q-q}) \right
         \}
-  \sin(\theta_h) \left\{
         g_{x,i}g_{z,i}\sin(\delta_{i-i})+ g_{x,q}g_{z,q}\sin(\delta_{q-q}) \right
         \} \right ] \cos(\Delta)  \right. \nonumber \\
&& \left. - \left [ \cos(\theta_h) \left\{
         g_{x,i}g_{z,q}\sin(\delta_{i-q})+ g_{x,q}g_{z,i}\sin(\delta_{q-i}) \right
         \}
+ \sin(\theta_h) \left\{
         g_{x,i}g_{z,q}\cos(\delta_{i-q})+ g_{x,q}g_{z,i}\cos(\delta_{q-i}) \right
         \} \right ] \sin(\Delta) \right \}.
\end{eqnarray}
\setlength{\arraycolsep}{5pt} \hrulefill
\vspace*{4pt}
\end{figure*}

Now, let us add all impairments such as phase/amplitude imbalance
and phase noise. The phase noise difference and sum between
downconverters for $X(t)$ and $Z(t)$ are defined by

\begin{align}
\varepsilon_{x-z}(t)=\varphi_x(t)-\varphi_z(t),\varepsilon_{x+z}(t)=\varphi_x(t)+\varphi_z(t).
\nonumber
\end{align}
Define the phase imbalance difference and sum of in-phase and
quadrature phase between downconverters for $X(t)$ and $Z(t)$ as
following

\begin{align}
\delta_{i-i}=\phi_{x,i}-\phi_{z,i},\delta_{q-q}=\phi_{x,q}-\phi_{z,q},
\nonumber \\
\delta_{i+i}=\phi_{x,i}+\phi_{z,i},\delta_{q+q}=\phi_{x,q}+\phi_{z,q},
\nonumber \\
\delta_{i-q}=\phi_{x,i}-\phi_{z,q},\delta_{q-i}=\phi_{x,q}-\phi_{z,i},
\nonumber \\
\delta_{i+q}=\phi_{x,i}+\phi_{z,q},\delta_{q+i}=\phi_{x,q}+\phi_{z,i}.\nonumber
\end{align}
Then, the echo channel estimation can be written as shown in
(\ref{ErrWPN}) where $\theta_x(t)$ is the angle of $X(t)$ assuming
uniformly distributed over $0$ to $2\pi$ and $\Delta$ is the phase
difference between downconverters for $Z(t)$ and $X(t)$. When in
majority of samples the sign of $Re\{ \tilde{X}^*(t) \tilde{Z}(t)
\}$ is equal to the sign of $|H|cos(\theta_h)$, the weight will be
updated to the right direction.

Due to random time varying phase from transmitted signal
$\theta_x(t)$, phase noise $\varphi_x(t)$ and $\varphi_z(t)$, the
echo channel estimation will fluctuate. However, by the low pass
filtering, those time varying noises will be filtered. Since we can
assume the phase noise and the phase of transmitted signal are
stationary, the integration over time can effectively perform
ensemble average. Assuming Gaussian phase noise, the ensemble
average over $X(t)$, $\varphi_x(t)$ and $\varphi_z(t)$ leads to
(\ref{EErrWPN}) where $\sigma^2$ is the variance of phase noise and
$P$ is the average power of $X(t)$. As can be seen, the phase noise
reduces the power of echo channel estimation. Hence, the performance
will be more sensitive to the noise. When all downconverters are
driven by the same oscillator, the phase noises can be identical. In
this scenario, the phase noise difference $\varepsilon_{x-z}(t)$
will be zero. Then, the scale factor $e^{-\sigma^2}$ disappears as
if no phase noise is present.

 \nocite{*}
\bibliographystyle{IEEE}

%

%
\begin{IEEEbiography}[{\includegraphics[width=1in,height=1.25in,clip,keepaspectratio]{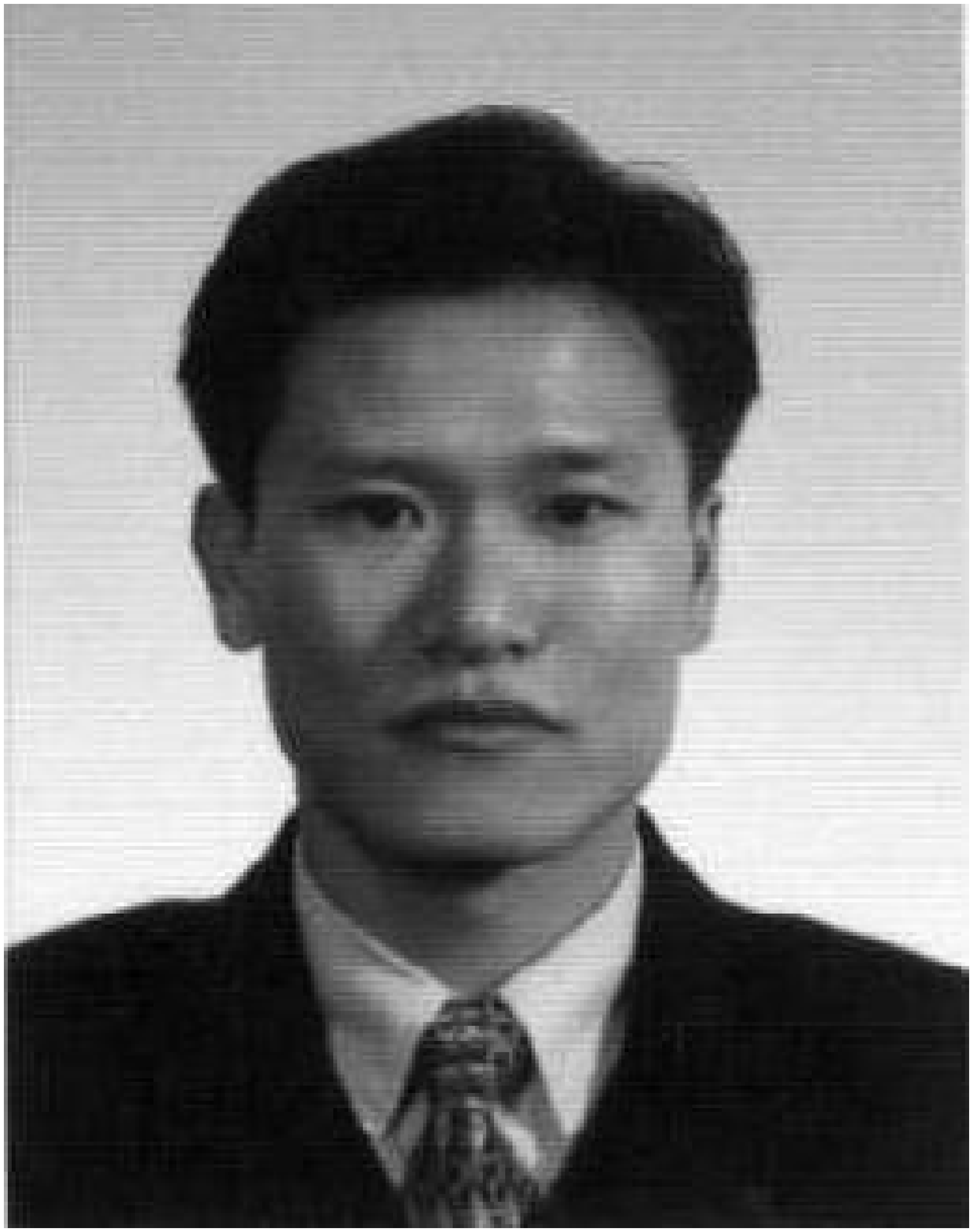}}]{Yang-Seok
Choi} (S'91,M'01) received the B.S. degree from Korea University,
Seoul, Korea, in 1990, the M.S.E.E. degree from Korea Advanced
Institute of Science and Technology (KAIST), Taejon, Korea, in 1992,
and the Ph.D. degree from Polytechnic University, Brooklyn, NY, USA
in 2000, all in electrical engineering. From 1992 to 1996, he was
with Samsung  Electronics, Co., Ltd., Suwon, Korea, where he
developed various modems for HDTV and DBS. During 2000 summer he
held a Summer intern position at AT\&T Labs-Research Shannon Lab,
Florham Park, NJ, USA. In 2000, he joined National Semiconductor,
East Brunswick, NJ, USA where he was involved in the development of
W-CDMA. During 2001-2002, he was a Senior Technical Staff Member at
AT\&T Labs-Research, Middletown, NJ, USA where he researched on MIMO
systems, OFDM systems and information theory. From 2002 to 2004 he
had been with ViVATO, Inc., Spokane, WA, USA working on MIMO OFDM
systems, smart antenna systems, and antenna/beam selection
techniques. In 2004, he joined Intel Corporation, Hillsboro, OR, USA
where he studied on broadband wireless communications systems and
was a director of Radio Systems Engineering leading Standards and
technology development. In 2013, he joined Intel Labs where he
researches future generation wireless systems. His research
interests include various aspects of wireless communications
systems.
\end{IEEEbiography}
%
%
\begin{IEEEbiography}[{\includegraphics[width=1in,height=1.25in,clip,keepaspectratio]{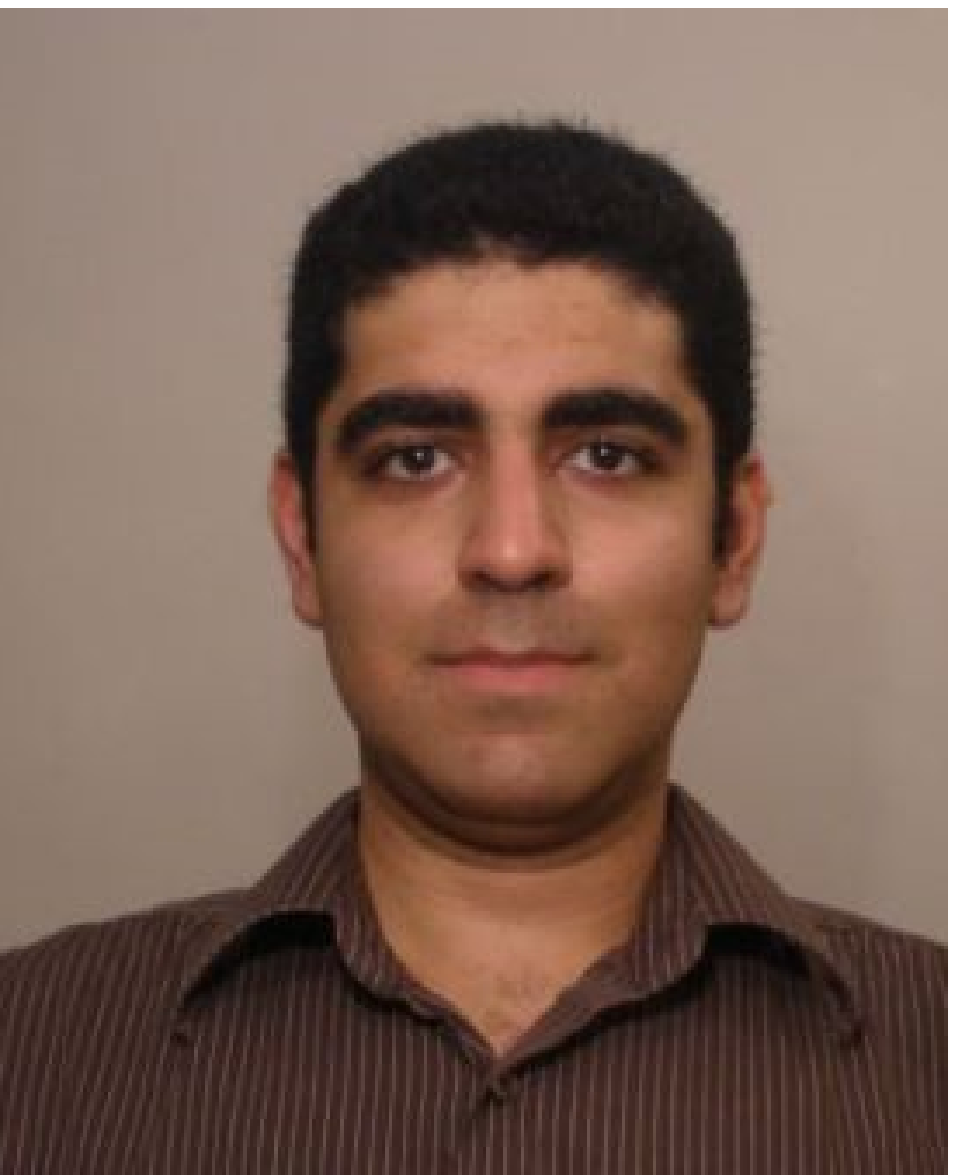}}]{Hooman
Shirani-Mehr} (S'03- M'10) received the B.S. degree from Sharif
University of Technology, Tehran, Iran, in 2001 and M.S. and Ph.D.
degrees from the University of Southern California, Los Angeles, CA
in 2006 and 2010, respectively, all in electrical engineering. Since
2010, he has been with Intel Corporation where he is currently
working on 3GPP LTE and LTE-Advanced wireless systems. His research
interests include communication theory, information theory and
signal processing with applications in wireless communications.
\end{IEEEbiography}

\end{document}